\documentclass[12pt]{article}

\usepackage{epsfig}
\begin{document}

\vspace*{2.0cm}

\begin{center}
{\large {\bf Properties of a relativistic equation of state \\
for collapse-driven supernovae} }

\vspace*{1.0cm}
K. Sumiyoshi$^{a,}$\footnote{e-mail: sumi@numazu-ct.ac.jp},
H. Suzuki$^{b,}$\footnote{e-mail: suzukih@ph.noda.tus.ac.jp},
S. Yamada$^{c,}$\footnote{Present address: Science and Engineering, 
Waseda University, e-mail: shoichi@heap.phys.waseda.ac.jp},
and H. Toki$^{d,}$\footnote{e-mail: toki@rcnp.osaka-u.ac.jp} \\
\vspace*{0.5cm}
      $^{a}$Numazu College of Technology, \\
            Ooka 3600, Numazu, Shizuoka 410-8501, Japan \\
      $^{b}$Faculty of Science and Technology, Tokyo University of Science, \\
            Yamazaki 2641, Noda, Chiba 278-8510, Japan \\
      $^{c}$Institute of Laser Engineering (ILE), Osaka University, \\
            Yamadaoka 2-6, Suita, Osaka 565-0871, Japan \\
      $^{d}$Research Center for Nuclear Physics (RCNP), Osaka University, \\
            Mihogaoka 10-1, Ibaraki, Osaka 567-0047, Japan \\
\end{center}

\vspace*{0.5cm}
\newpage
%%%%%%%%%%%%%%%%%%%%%%%%%%%%%%%%%%%%%%%%%%%%%%%%%%%%%%%%%%%%
\begin{abstract}
We study characteristics of the relativistic equation of state (EOS)
for collapse-driven supernovae, 
which is derived by relativistic nuclear many body theory.  
Recently the relativistic EOS table has become available 
as a new complete set of physical EOS for numerical simulations 
of supernova explosion.  
We examine this EOS table by using general relativistic hydrodynamics 
of the gravitational collapse and bounce of supernova cores.  
In order to study dense matter in dynamical situation, 
we perform simplified calculations of core collapse and bounce 
by following adiabatic collapse with the fixed electron fraction 
for a series of progenitor models.  
This is intended to 
give us ``approximate models'' of prompt explosion.  
We investigate the profiles of thermodynamical quantities 
and the compositions during collapse and bounce.  
We also perform the calculations with the Lattimer-Swesty EOS 
to compare the properties of dense matter.
As a measure of the stiffness of the EOS, 
we examine the explosion energy 
of the prompt explosion with electron capture totally suppressed.  
We study the derivative of the thermodynamical quantities 
obtained by the relativistic EOS to 
discuss the convective condition in neutron-rich environment, 
which may be important in the delayed explosion.  
\end{abstract}

%%%%%%%%%%%%%%%%%%%%%%%%%%%%%%%%%%%%%%%%%%%%%%%%%%%%%%%%%%%%
\newpage
\section{Introduction}

Clarifying the mechanism of core-collapse supernova explosion 
is fascinating.  
It is a challenging problem which demands extensive research 
efforts in physics and astrophysics.  
Towards the final answer, one has to conduct sophisticated 
numerical simulations treating all ingredients of microphysics 
and macrophysics.  
Recently, numerical simulations solving hydrodynamics 
together with the Boltzmann equation for the 
neutrino transfer have become available 
in spherical symmetry \cite{Mez93, Yam99, Bur00, Ram00, Mez01, Lie01}.
In these simulations, the basic equations of hydrodynamics 
and neutrino transfer have been solved directly 
together with the careful implementation of microphysics.  
This recent progress, removing one of uncertainties due to 
approximate neutrino transfer, 
has shed light again on the importance of the microphysics 
such as neutrino interactions and properties of dense matter.  

An important microphysics ingredient for supernova simulations is 
the equation of state (EOS) of dense matter.  
It determines the stellar structure, the hydrodynamics 
and the reaction rates through the determination of pressure, 
entropy and chemical compositions.
Although study of dense matter for supernova research apparently has a long 
history, there are only a few studies to cover 
the whole range of density, electron fraction and temperature 
in supernova environment.  
Because one has to simulate the whole phenomena starting 
from collapsing iron cores to cooling neutron stars, 
it is necessary to provide thermodynamical quantities in a 
wide range of density, composition and temperature 
in a usable and complete form for numerical simulations.  
Such efforts to provide the EOS 
have been done so far by a limited number of groups.  
The EOS formulated in terms of nuclear 
parameters by Baron, Cooperstein and Kahana \cite{Bar85} 
has been used to study the influence of the 
EOS on the explosion.  
Hillebrandt and Wolff \cite{Hil85}
have taken the Skyrme Hartree-Fock approach 
to provide the data table of EOS for 
supernova simulations.  
This EOS table has been used in some simulations 
\cite{Hil85, Suz90, Suz93},
but is not currently used.  
Lattimer and Swesty \cite{Lat91} 
have utilized the compressible liquid-drop model 
to provide the EOS as a numerical routine for supernova simulations.  
This EOS has been used in many simulations 
of supernova explosion these years.  
However, 
it has been difficult to assess the dependence 
of the supernova phenomena on the EOS 
since available EOS's 
based on different nuclear models are limited 
(See, however, \cite{Bru86, Bru89a, Bru89b, Swe94}).  
% previous studies of EOS effect on supernova

Recently, a new complete EOS 
for supernova simulations has become available \cite{She98a, She98b}.  
The relativistic mean field (RMF) theory with a Thomas-Fermi approach 
has been applied to the derivation of the supernova EOS.  
The RMF theory has been successful 
to reproduce the saturation properties, masses and radii 
of nuclei, and proton-nucleus scattering data 
\cite{Ser86}.
The effective interaction used in the RMF theory is checked 
by the recent experimental data of unstable nuclei 
in neutron-rich environment close to 
astrophysical situations.  

We stress that the RMF theory is based 
on the relativistic Br\"uckner-Hartree-Fock (RBHF) theory 
\cite{Bro90}.
The RBHF theory, which is a 
microscopic and relativistic many body theory, 
has been shown to be successful to reproduce the 
saturation of nuclear matter starting from the 
nucleon-nucleon interactions determined by the 
scattering experiments.  
This is in good contrast with non-relativistic 
many body frameworks which can account for the saturation 
only with the introduction of extra three-body 
interactions.  
The effective lagrangian of the RMF theory 
has been extended to take account of the 
behavior of the RBHF theory \cite{Sug94}.  
It is also noteworthy that relativistic frameworks 
automatically satisfy the causality, which is 
the condition that the sound velocity should not 
exceed the light velocity, while the causality is 
often violated in non-relativistic frameworks.  

Having a new set of EOS table, 
numerical simulations of hydrodynamics with neutrino transfer 
are awaited to investigate the influence of the relativistic EOS 
on the supernova dynamics as well as supernova neutrinos.  
Before we proceed to such elaborate numerical simulations,  
which require large computational efforts, 
we would like to examine the properties of the relativistic EOS 
in dynamical situation during supernova explosion 
which may not be apparent in a common practice of calculations 
of neutron star structure.  
It will also be helpful to provide the basic information 
on the behavior of relativistic EOS during collapse and bounce 
for forthcoming detailed simulations.  
For this purpose, we perform hydrodynamical calculations 
of adiabatic collapse with the relativistic EOS table 
dropping off the treatment of neutrino transfer.  
We adopt presupernova models by Woosley-Weaver \cite{Woo95}
and Nomoto et al. \cite{Nom97} as realistic initial models.  
We fix the electron fraction at the initial value, 
which maximizes the shock energy at bounce, 
as working models of explosion.  
With these ``trial models'', we examine the properties of dense matter 
such as compositions obtained from our EOS 
during core collapse and bounce.  
We evaluate the energy of shock at breakout as one of measures 
of the EOS for a series of presupernova models.  
For comparison, we perform the hydrodynamical calculations 
with the Lattimer-Swesty EOS 
and explore the difference of the properties of dense matter 
and the shock energy.  

In addition to getting the information mentioned above, 
we would like to test whether the relativistic 
EOS table works well in numerical simulations during 
core collapse and explosion.  
This is because some EOS tables so far have been 
causing troubles in numerical simulations due to 
lack or inconsistency of data.  
The relativistic EOS table has been calculated 
in a wide regime of environment 
(density, proton fraction and temperature)
to provide 
all necessary quantities of dense matter, 
which are tabulated for numerical simulations \cite{She98b}
of various astrophysical 
phenomena such as supernovae and neutron star mergers.  

This paper is arranged as follows.  
In section 2, after a brief introduction of 
collapse-driven supernova explosions, 
we explain our hydrodynamical calculations 
which are used for the study of the relativistic EOS.  
In section 3, we present our numerical results.  
We first show the initial profiles of presupernova core 
derived with the relativistic EOS table (section 3.1).  
In section 3.2, we display the profiles of thermodynamical quantities 
from the relativistic EOS during collapse and bounce.  
We compare the results with the case using the Lattimer-Swesty EOS. 
In section 4, 
we also discuss the effect of the relativistic EOS 
on the Ledoux criterion of convection.  
Summary will be given in section 5.  

%%%%%%%%%%%%%%%%%%%%%%%%%%%%%%%%%%%%%%%%%%%%%%%%%%%%%%%%%%%%
\section{Hydrodynamical simulation}

\subsection{Treatment of core-collapse and explosion}

The collapse-driven supernova explosion originates 
from massive stars with $M \geq 10M_{\odot}$ 
(for reviews, see \cite{Bet90, Suz94, Jan01}, for example).  
At the end of the thermonuclear stellar evolutions, 
the iron core is formed at the center of a presupernova star.  
This iron core becomes gravitationally unstable and starts collapsing 
due to the 
photo-dissociation and electron captures, which reduce 
the pressure to support the core.  
The gravitational collapse proceeds 
homologously in the inner part of the iron core 
and supersonically in the outer part of the iron core.  
When the central density exceeds the nuclear 
matter density, the inner core bounces back 
due to a hard core component of nuclear force.  
Due to this bounce, a shock wave is formed and 
it propagates outward.  
The shock wave goes through the outer core region, consuming 
its energy to dissociate irons into nucleons and 
to push back the falling material against the gravitational 
attraction.  

If the shock wave reaches the surface of the iron 
core without stalling on the way, it 
blows off the mantle outside the iron core 
to cause a supernova explosion.  
This simple scenario of hydrodynamical explosion 
is called ``prompt explosion''.  
If the shock wave stalls inside the outer core 
region, some extra mechanism is necessary 
to revive the stalled shock.  
For example, the neutrino heating behind the shock 
might save 
the failed shock wave and give a successful supernova 
explosion in a relatively long time scale around 100 msec.  
This type of supernova mechanism is called 
``delayed explosion''.  

Although prompt explosion is a simple scenario, 
it is generally believed 
that the explosion does not occur in this way 
for a major fraction of progenitors 
(see recent studies in \cite{Ram00, Mez01, Lie01}).  
In the past, Baron et al. have found that the prompt 
explosion occurs for the combination of soft EOS's 
and general relativistic hydrodynamics \cite{Bar85}.
However, later studies by Myra and Bludman \cite{Myr89} 
have shown that those explosions obtained in an approximate 
leakage scheme do not occur in a better treatment 
of neutrino transport (multigroup flux-limited diffusion 
approximation) with an appropriate treatment of neutrino interactions 
such as neutrino-electron scattering.  
Further systematic investigations 
by Bruenn and Swesty et al. \cite{Bru89a, Bru89b, Swe94}
have confirmed that no prompt explosion occurs except for 
extreme range of parameters of EOS and 
presupernova stars.  
Because of the difficulty of prompt explosion, 
especially for massive stars such as SN1987A, 
the delayed mechanism is believed to be the cause of 
the explosion.  
However, detailed processes to revive 
the stalled shock have been under debates \cite{Jan01}.  

The EOS plays essential roles in supernova 
dynamics both in direct and indirect manners.  
A softer EOS gives more compact cores at bounce 
with larger binding energy 
and provides larger energy to shock wave.  
The dependence of supernova explosion on the softness 
of EOS has been studied by changing 
incompressibility of nuclear matter \cite{Bar85, Bru89b, Swe94} 
(See also \cite{Tak88} in a different context).
Smaller incompressibility is indeed found to be preferable.

The composition of dense matter is also another 
important factor of explosion.  
The abundances of proton, neutron, alpha particle 
and nucleus determine the reaction rates 
of electron captures, neutrino scatterings and others.
The species of nucleus in dense matter 
is essential for evaluation of the electron capture rate 
on nuclei.  
When the electron captures on nuclei are suppressed 
in the high density regime, 
the free proton fraction is important to determine 
the electron capture rate on protons 
(see, however, \cite{Lan03} on the importance 
of electron captures on nuclei).  
Those electron capture rates control the neutrino emissivity 
during the collapse, resultant trapped lepton fraction 
and the size of inner core.  
A larger inner core leads to a larger initial shock energy 
and a smaller outer core through which the shock wave 
must propagate.  
Therefore, it is preferable for shock propagations 
if the EOS 
provides the composition which suppresses electron captures.  
For the determination of composition, the description of 
nuclear interaction for nuclei and matter is essential.  
As for the free proton fraction, it is closely related to 
the symmetry energy.  
The influence of the free proton fraction on supernova 
explosion has been studied \cite{Bru89a, Swe94}.  
It has been found that larger symmetry energies provide
smaller free proton fractions, which lead to 
a larger electron fraction.  
% more discussion of works by Bruenn and Swesty
% symmetry energy value in RMF and nuclear data

In the current study, we investigate the properties 
of the relativistic EOS using the collapse and bounce 
of the supernova cores.  
Since the new EOS table is derived in the relativistic many 
body framework unlike the previous EOS sets 
for supernova simulations, we would like to know 
the basic properties of the new EOS in a 
dynamical context.  
Since the current EOS table is based on the data of 
unstable nuclei and the symmetry energy part of nuclear 
interactions is better constrained, the composition of 
dense matter may be different from those in the previous EOS sets. 
Since we employ simplified adiabatic collapse with no neutrino 
transfer, 
we do not intend to discuss the supernova mechanism itself.  
Instead, the current study will provide the basic information 
of the relativistic EOS table in the physical situations 
close to the realistic supernova explosion 
and will serve as a guide to proceed to more sophisticated 
numerical simulations.  
% As a reference,
% we make comparison of properties of dense matter 
% with those in the EOS by Lattimer and Swesty.  
% We evaluate the shock energies for different presupernova 
% models as one of measures for the EOS stiffness.  
% More extensive study by general relativistic hydrodynamics 
% with Boltzmann solver of neutrino transfer, which is 
% extremely time-consuming, is currently in preparation 
% and will be published elsewhere.  

%\subsection{Numerical code}

We adopt the implicit numerical code for the general relativistic 
and spherically symmetric hydrodynamics \cite{Yam97}.
The implicit time differencing is advantageous 
to carry the hydrodynamical calculations for a long time.  
The code uses a Lagrangian baryon mass coordinate 
(baryon mass enclosed within a certain radius) 
as a radial coordinate.  
We refer to the article by Yamada \cite{Yam97}
for details of 
the numerical treatment of hydrodynamics.  
% shock capture
% Simple numerical tests of hydrodynamics have been already 
% done to check the applicability of the numerical code.  
This numerical code is specifically designed for the study of 
supernova explosion to treat both the hydrodynamics and the 
neutrino transfer.  
The Boltzmann solver of neutrino transfer is recently 
implemented in the numerical code \cite{Yam99}.
Detailed simulations of gravitational core collapse 
based on this fully implicit and general relativistic 
code with neutrino transfer will be presented elsewhere.  

%Meanwhile, in the current study, we perform the calculations of 
%pure hydrodynamics of core collapse without neutrino transfer 
%before we proceed to the full simulation.  
%We would like to employ the results as ``model hydrodynamics'' 
%of core collapse, bounce and model explosion 
%in order to assess the relativistic EOS table in the dynamical context.
%For this particular purpose, we also drop off 
%electron captures and freeze the evolution of the electron 
%fraction.  
%Because of high values of the fixed electron fraction, 
%the size of inner core is large and the explosion energy 
%is maximized.  
%The heating and cooling via neutrinos are also dropped.  
%Therefore, we study adiabatic collapse.
%That is, the entropy per baryon of each mass element 
%remains constant except for the increase due to the shock.  

\subsection{Equation of state of dense matter}

We adopt the table of the relativistic EOS, which 
is recently derived for supernova simulations 
\cite{She98a, She98b} 
in the relativistic nuclear many body framework.  
Based on the relativistic Br\"uckner-Hartree-Fock theory, 
which is successful to reproduce the nuclear matter 
saturation, the relativistic mean field (RMF) theory is 
constructed to describe nuclear matter and nuclei 
\cite{Sug94}.
The same RMF framework has been applied to the 
systematic study of about 2000 nuclei, including 
deformed ones, up to the drip line in the nuclear chart 
and the nuclear data such as masses and radii are 
successfully reproduced \cite{Hir97}.
Having thus being checked by the data of stable and 
unstable nuclei, the RMF framework has been used to 
study the properties of dense matter in supernovae 
\cite{Sum95b}.
The RMF framework with the parameter set TM1, 
which was determined as the best one to reproduce 
the properties of finite nuclei 
\cite{Sug94},
provides the uniform nuclear matter with 
the incompressibility of 281 MeV 
and the symmetry energy of 36.9 MeV at the saturation 
density.  
We remark that the behavior of nuclear matter at 
high density in the RMF framework is similar to 
the one in the relativistic Br\"uckner-Hartree-Fock theory 
due to the inclusion of the non-linear terms in the 
RMF lagrangian.  
The maximum neutron star mass calculated for the cold 
neutron star matter in the RMF with TM1 is 2.2 $M_{\odot}$.  
The RMF framework has been extended 
with the Thomas-Fermi approximation 
to describe not only the homogenous but also inhomogeneous matter 
in supernovae, 
which contains neutrons, protons, alpha particles 
and heavy nuclei as a representative species.  
The properties of dense matter (energy, pressure, 
entropy, chemical potential and so on)
at various densities, 
proton fractions and temperatures are calculated 
to construct the numerical data table for simulations.  
The table covers the wide range of density 
($10^{5.1} \sim 10^{15.4}$ g/cm$^{3}$), 
electron fraction 
($0.0 \sim 0.56$), 
and temperature 
($0 \sim 100$ MeV), 
which is required for supernova simulations.  
The relativistic EOS table is available upon 
request to one of the authors, K. Sumiyoshi.  
The electron/positron and photon contributions 
as non-interacting particles 
are added to the nuclear contribution of the EOS.  
% The arbitrary degeneracy of electrons and 
% disappearance of positrons at low temperatures 
% are properly treated.  
The relativistic EOS table has been applied to 
numerical simulations of r-process in 
the neutrino-driven winds \cite{Sum00} and 
in supernova explosions \cite{Sum01}.
A proto-type of the EOS table has been used in 
the numerical simulations of the proto-neutron 
star cooling \cite{Sum95c}.
% cf. Lattimer-Swesty, Wolff

For comparison, we adopt also the EOS by Lattimer and Swesty \cite{Lat91}, 
which has been used for recent supernova simulations.  
The EOS is based on the compressible liquid drop model for 
nuclei together with dripped nucleons \cite{Lat85}.  
The bulk energy of nuclear matter is expressed in terms 
of density, proton fraction and temperature 
with nuclear parameters.
The values of nuclear parameters are chosen to be the ones 
suggested from nuclear mass formulae and other theoretical 
studies mostly with the Skyrme I$'$ force.  
Among various parameters, the symmetry energy is set to be 29.3 MeV, 
which is smaller than the value in the relativistic EOS.  
As for the incompressibility, there are three choices 
with different values of 180, 220, 375 MeV.  
We use, for comparison, the EOS with 180 MeV, 
which has been often used for supernova studies.  

\subsection{Initial models}

We adopt the presupernova models of massive stars provided by 
Woosley (WW95) \cite{Woo95, Woo99}
and by Nomoto (N97) \cite{Nom97, Nom00} (see also \cite{Nom88}).
From WW95, 
we have used 11, 12, 13, 15, 18 and 20 $M_{\odot}$ models 
among the models ranging from 11$M_{\odot}$ to 40$M_{\odot}$.  
From N97, we have used the models with Helium cores of 
3.3, 4, 5 and 6 $M_{\odot}$, which correspond to the progenitor mass of 
13, 15, 18 and 20 $M_{\odot}$, respectively.  
In Table 1, we list the models we have used for the calculations.  
We utilize the central core part which covers the Fe core and 
the outer layer with 
the density down to about 10$^{6}$ g/cm$^{3}$.  
The masses of the calculated central core and the Fe core 
are listed also in Table 1.  % metalicity
We take density, electron fraction, temperature 
and radius as a function of baryon mass coordinate 
from the original data and derive other quantities 
using the relativistic EOS table.  
The number of the baryon mass grid points is 100.  
The grid size decreases linearly from the center 
to the surface to provide enough grid points around the 
surface.  
As for the outer boundary condition, we fixed 
density, internal energy, entropy and electron fraction 
of the most outer grid point at the original values 
in the initial model.  
The original configuration of the central core 
of the presupernova models is already marginally unstable 
to gravitational collapse.  
When we start the numerical calculation with the constructed 
initial configuration, the inner core starts collapsing 
immediately.  
We have checked that, if we increase slightly the electron 
fraction in the original models, the core remains stable 
and does not start collapsing.  
Therefore, the profile of electron fraction in the 
models has just marginal values to trigger the gravitational 
collapse.  

%%%%%%%%%%%%%%%%%%%%%%%%%%%%%%%%%%%%%%%%%%%%%%%%%%%%%%%%%%%%
\section{Numerical results}

\subsection{Initial Composition}

Before we proceed to the results of hydrodynamics, 
we discuss the composition of the central core 
derived with the relativistic EOS table.  
The composition just prior to the collapse is 
interesting since it affects the trapped lepton 
fraction through weak interactions.  
The trapped lepton fraction in turn determines 
the size of the inner core and the available 
explosion energy.  
The species of nuclei in the central core 
are important to evaluate the electron capture 
rates on nuclei in the hot and dense matter.  
After the electron capture on nuclei is blocked
due to the nuclear shell effects, 
the electron capture on free protons becomes 
important.  
Therefore, the mass fraction of free protons 
is also important quantity to determine the evolution 
of electron (lepton) fraction 
during the collapse stage \cite{Bet90, Suz94}.  

%, neutron 
Figure \ref{fig:initial} displays 
the mass fraction of particles (upper panel), 
the proton, neutron and mass numbers of nuclei (lower panel) 
as a function of baryon mass coordinate 
in the central core of 15$M_{\odot}$ model of WW95.  
The values with the relativistic EOS and the Lattimer-Swesty 
EOS \cite{Lat91} are shown by thick and thin lines, respectively.
The quantities of the Lattimer-Swesty EOS are obtained 
by giving the same profiles of density, temperature 
and electron fraction 
(See Fig. \ref{fig:density}
for the initial profile of density).
% Alpha difference

In the upper panel, the initial profile of 
electron fraction is also shown.  
The central value of electron fraction in the initial 
model is 0.42.
It can be seen that the mass fractions 
are substantially different between the relativistic EOS and 
the Lattimer-Swesty EOS.
The mass fractions of alpha particle and neutron 
in the relativistic EOS are larger than those 
in the Lattimer-Swesty EOS.  
The free proton fraction is about one order of 
magnitude smaller than the numbers obtained 
with the Lattimer-Swesty EOS.  
Having a smaller fraction of free proton is favorable 
for the explosion since the electron capture 
is largely reduced, leading to a stronger prompt-shock \cite{Bru89a}.
% DISCUSS MORE ON BRUENN'S RESULT

The reduction of free proton fraction comes 
from an effect of the large symmetry energy.  
A larger symmetry energy provides a 
larger difference between chemical potentials 
of protons and neutrons.  
A lower chemical potential for protons leads to a smaller 
free proton fraction and a higher chemical potential 
for neutrons gives a larger free neutron fraction.  
Note that the symmetry energy 
of the relativistic EOS (36.9 MeV) is larger 
than that 
of the Lattimer-Swesty EOS (29.3 MeV).
The value of the symmetry energy is constrained 
mainly by nuclear masses, but there is still an 
allowable range.  
Systematic measurements of the radii of neutron-rich 
nuclei may further constrain the size of the symmetry energy 
since the thickness of neutron skins is sensitive 
to the symmetry energy.  
We stress that 
the symmetry energy of the relativistic EOS 
has been checked by the nuclear structure 
calculations of the radii of neutron-rich nuclei 
in the RMF framework \cite{Sug94, Sug96} together with 
the recent experimental data of neutron skins 
\cite{Suz95, Oza01}.  

It has been known that the relativistic many body 
frameworks provide a larger effect of the symmetry 
energy on nuclei and uniform nuclear matter 
than the non-relativistic many body frameworks 
\cite{Sum93}.
The relativistic many body frameworks provide 
stronger density dependence of the symmetry energy.
Accordingly, 
the proton fraction in the cold neutron star matter 
in the relativistic many body frameworks is larger 
than that in the non-relativistic frameworks \cite{Sum95a}.  
A large proton fraction can lead to the rapid cooling of neutron 
stars due to the direct URCA process \cite{Lat91a}.  
At the same time, the thickness of neutron skins of 
neutron-rich nuclei is enhanced in the relativistic 
frameworks.  
It has been shown that 
the systematics of neutron skins of isotopes 
and the EOS of asymmetric matter are related to 
each other and the behavior of asymmetric matter 
can be probed by systematic measurements 
of unstable nuclei \cite{Oya98}.
A precise measurement of neutron radius of $^{208}$Pb 
in future would provide more information on the EOS of 
asymmetric matter \cite{Bro00, Hor02}.  
It is interesting to see if 
these differences of the nuclear properties 
coming from the new relativistic frameworks 
have influence on the supernova 
explosion through the change of compositions 
and reaction rates.  

It is remarkable that heavy nuclei beyond 
A=60 appear in a wide region of the core.  
It is also interesting to see that the neutron 
number exceeds 40 in the central region, 
where electron capture might be suppressed 
according to the shell blocking 
\cite{Ful82, Bru85} (see, however, \cite{Lan03}).
Compared with the results in the Lattimer-Swesty EOS,
the species of nuclei is about 10\% heavier.  
The slightly larger value of binding energy 
of nuclear matter in the relativistic EOS table (B=16.3 MeV) 
than in the Lattimer-Swesty EOS (B=16.0 MeV), and/or 
the different treatments of nuclear surface, neutron skin 
and translational 
energy affect the determination of species, which should be 
studied further carefully.  
These findings of heavier nuclei may suggest that the detailed study 
of electron capture rates in A=60 region and beyond 
is important for the outcome of 
the core collapse.  
In recent shell-model calculations  
\cite{Cau99, Lan00}
of weak-interaction 
rates for nuclei in the mass range A=45-65 in the 
stellar environment, electron capture rates are 
significantly smaller than the standard set of 
Fuller, Fowler and Newman 
\cite{Ful80, Ful82a, Ful82b, Ful85}.
Calculations of presupernova evolution of 
massive stars with improved rates for weak interactions 
\cite{Heg00}
provide larger electron fractions, 
leading to a possible increase of explosion energy.  
This trend of suppression of electron capture on 
nuclei might be a key toward successful explosions and 
it strengthens the motivation to study core collapse in detail 
by the improved EOS and weak interactions.  
% More heavier nuclei

One has to be cautious, however, that there is a negative 
feedback in the deleptonization during collapse.  
If the electron fraction is large in the initial stage, 
the free proton fraction becomes large and electron captures 
on free protons drives back the electron fraction small.  
Resulting trapping lepton fractions turn out similar 
even from the different electron fractions in progenitors 
\cite{Lie02}.  
This is because the free proton fraction depends steeply 
on the electron fraction.  
For example, the free proton fraction becomes 
smaller by a factor of 10 if we decrease the electron 
fraction by $\Delta Y_e=0.03$ from the value ($Y_e=0.42$) 
at the center of initial model.
We have looked into the case of the Lattimer-Swesty 
EOS and have found that 
the dependence of the free proton fraction on 
the electron fraction is similar in this region.
To get the same value with the Lattimer-Swesty EOS 
for the free proton fraction, 
a larger value by $\Delta Y_e \sim 0.02$ can be
possible in the relativistic EOS.
If the electron captures would proceed up to a 
certain value of the free proton fraction, 
the resulting lepton fraction might be larger 
by about 0.02, which might lead to a slightly 
larger inner core, in the relativistic EOS.
Whether this difference of the free proton 
fraction influences the core collapse remains 
to be seen in more detailed calculations.

We note here that the current comparison is made by adopting 
the temperature profiles from the presupernova models 
instead of the entropy profiles.  
One has to be careful about the choice of thermodynamical 
inputs (temperature or entropy) to discuss the composition 
of matter since the composition is sensitive to temperature.  
Extensive comparisons with the Lattimer-Swesty EOS during 
the core collapse and bounce are currently being made and 
will be published elsewhere.  

\subsection{Profiles during core collapse and bounce}

\begin{figure}
\begin{center}
\epsfig{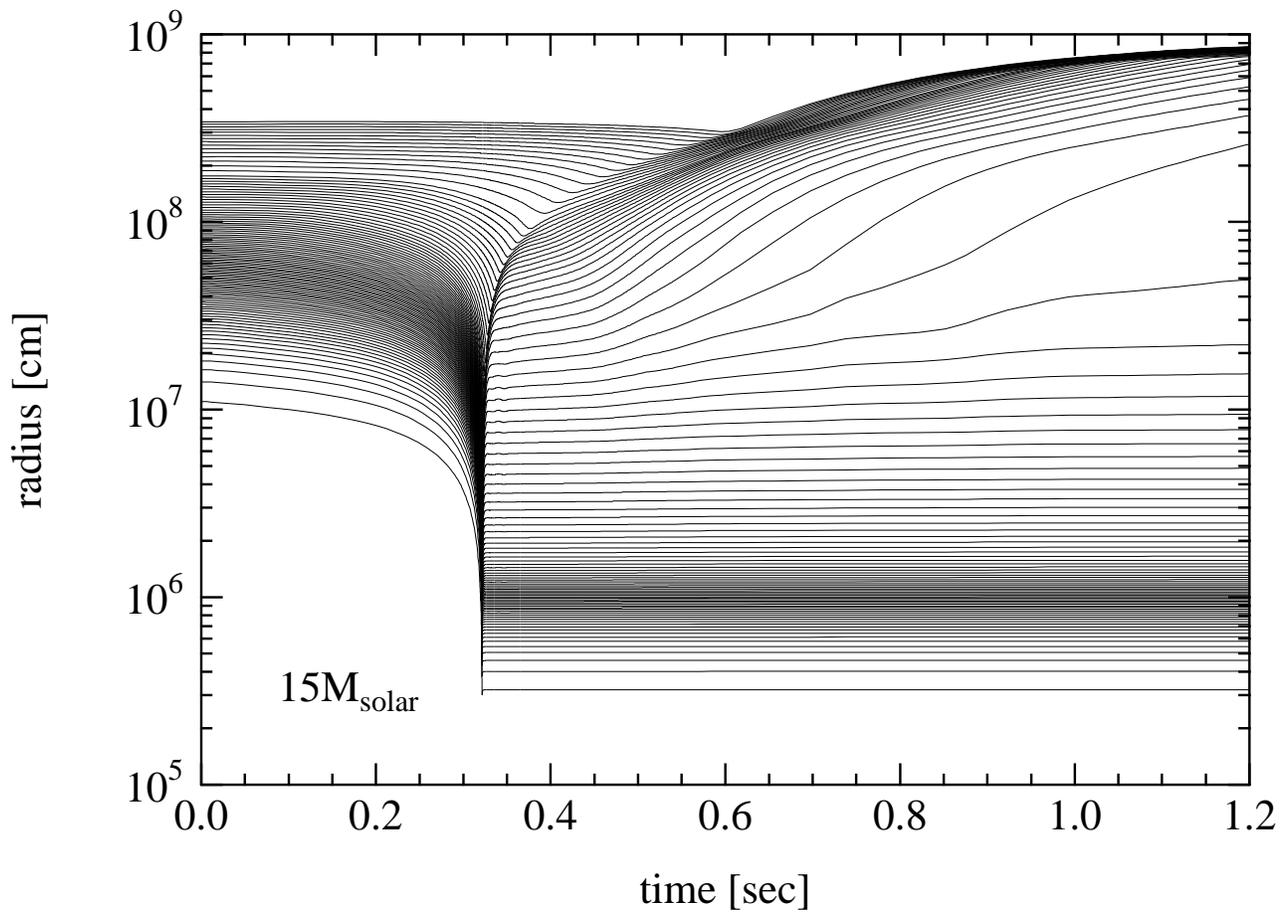}
\end{center}
\caption{The trajectories of mass mesh in radius are displayed 
as a function of time in hydrodynamical calculation of 
15 $M_{\odot}$ models of WW95.}
\label{fig:vw15ms04}
\end{figure}

\begin{figure}
\begin{center}
\epsfig{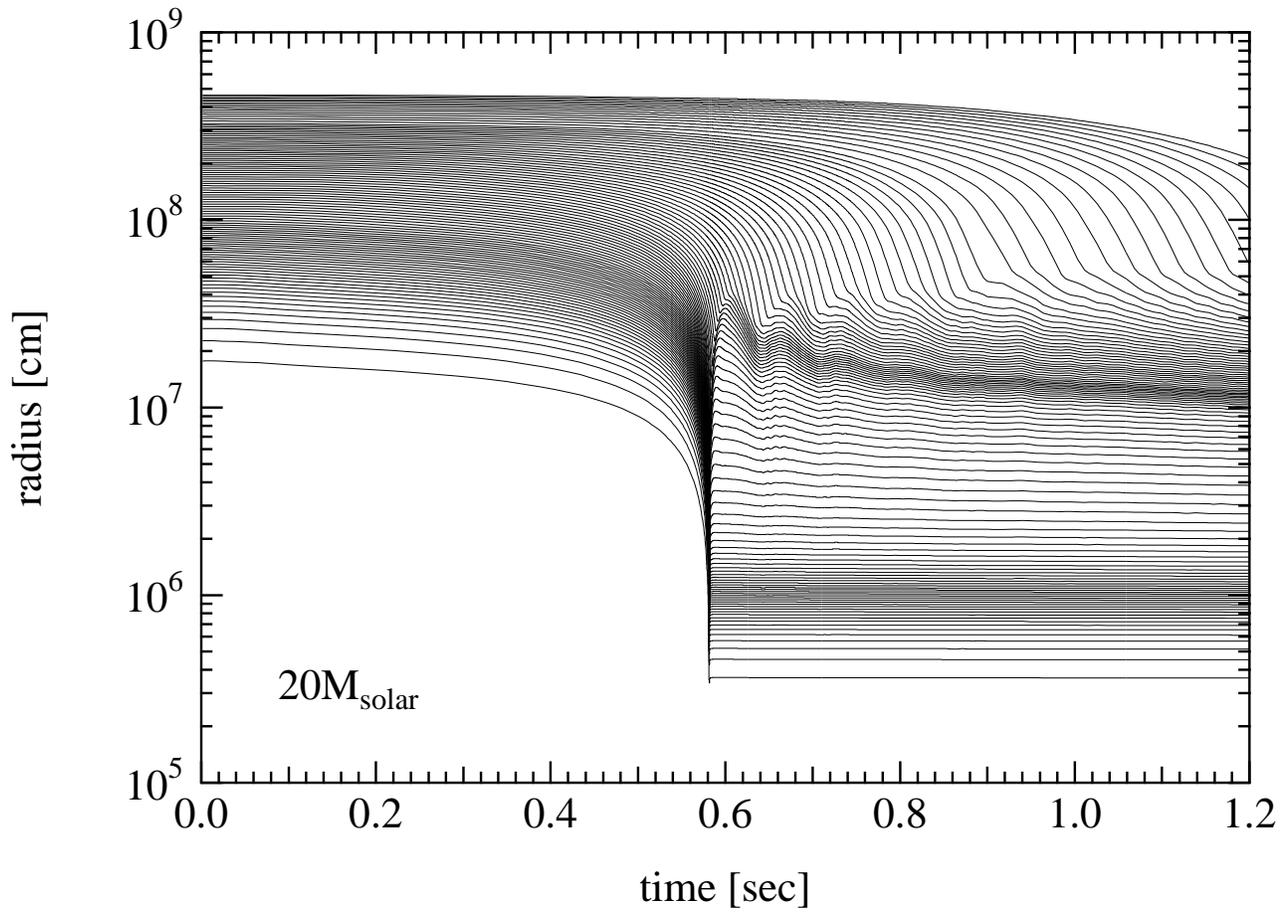}
\end{center}
\caption{The trajectories of mass mesh in radius are displayed 
as a function of time in hydrodynamical calculation of 
20 $M_{\odot}$ models of WW95.}
\label{fig:vw20ms01}
\end{figure}

Figures \ref{fig:vw15ms04} and \ref{fig:vw20ms01} 
demonstrate successful and unsuccessful explosions 
in hydrodynamical calculations 
starting from the initial configurations of the 
15 and 20 $M_{\odot}$ presupernova models of WW95.  
In the cases of 11, 12 and 15 $M_{\odot}$ models, 
a shock wave launched at the core bounce goes through the outer 
core, leading to a successful prompt explosion.  
On the other hand, in the cases of 13, 18 and 20 $M_{\odot}$ 
models, a shock wave stalls after the bounce due to 
photo-dissociation and infalling material.  
We list up the size of iron core in Table 1.  
It is natural to see that the success of prompt explosion 
is clearly correlated with the size of iron core.  
The size of iron core is small (1.32$M_{\odot}$) in all the 
successful cases, whereas it exceeds 1.4$M_{\odot}$ 
in the unsuccessful cases.  
Interestingly, the 13 $M_{\odot}$ model has an iron core of 
1.41 $M_{\odot}$, being different from neighboring 11 and 
15 $M_{\odot}$ models and does not show a prompt explosion.  
In the cases of N97 models, the explosion is successful 
for 13, 15 and 18 $M_{\odot}$.
The size of iron core is again small (1.18, 1.28 and 1.36 $M_{\odot}$ 
for 13, 15 and 18 $M_{\odot}$ models, respectively) in these 
cases \cite{Nom97}.
The case of 20 $M_{\odot}$ model does not show a prompt 
explosion.  
The size of iron core in this case is 1.40 $M_{\odot}$, 
which suggests that the threshold is around $1.4 M_{\odot}$.  

Explosion energy amounts to 1.7 foe (foe = 10$^{51}$ erg) 
for 11 and 12 $M_{\odot}$ models and 1.5 foe for 15 $M_{\odot}$ 
model of WW95, whereas 1.8, 1.7 and 1.7 foe for 13, 
15 and 18 $M_{\odot}$ models of N97, respectively.  
Here we define that the explosion energy is the energy of 
ejecta (gravitational mass minus baryon mass of ejecta) 
in the general relativistic formulation (see Appendix and \cite{Yam97}).  
We list up the explosion energy and the remnant mass 
for the cases of successful explosion in Table 1.  
These results confirm the correspondence between the iron 
core mass (not progenitor mass) and explosion energy.  
It has been argued that a smaller iron core mass results 
in a larger explosion energy since the amount of dissociation 
of iron nuclei becomes smaller for the similar size of unshocked 
cores \cite{Arn82} (see also \cite{Suz94}).
We note here again that the electron fraction is fixed 
at the initial value, which maximizes the shock energy at bounce, 
in order to obtain working models of explosion.  
The explosion energy evaluated here is the maximum value 
attained with the relativistic EOS table and 
should be regarded as one of measures 
of the stiffness of EOS.

In more realistic simulations with the neutrino transfer, 
the shock energy is lost also 
by the cooling due to neutrino emissions and 
the pressure by falling material 
in addition to the dissociation of iron nuclei.  
Whether the stalled shock wave revives or not depends 
probably on the neutrino heating mechanism and 
the explosion energy may depend more on how the delayed 
explosion occurs.  
% Ejected mass? r-process
% Further careful study of stellar evolution prior 
% to the collapse is important.  

\begin{figure}
\begin{center}
\epsfig{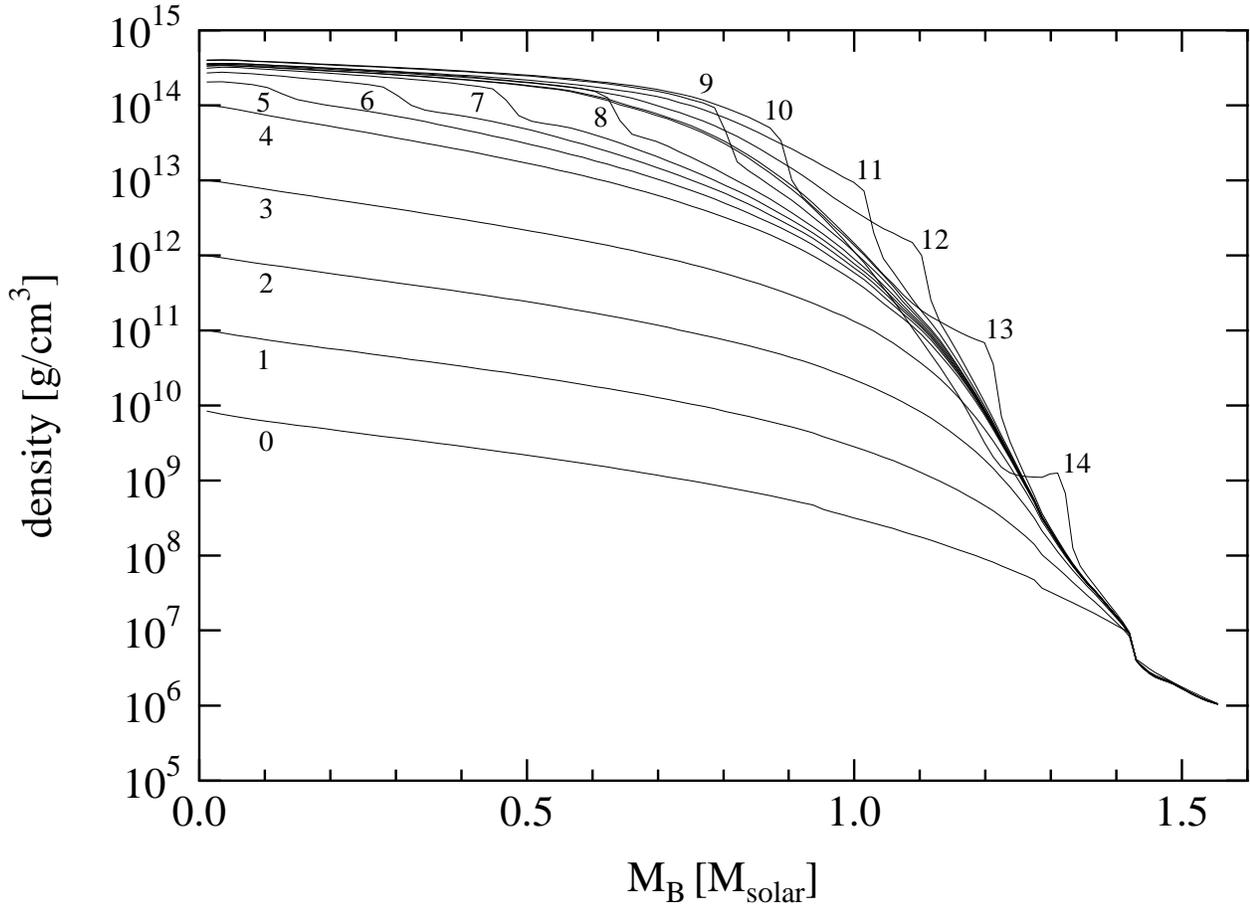}
\end{center}
\caption{The density profiles at selected times are shown as a function 
of baryon mass coordinate for the case of 15 $M_{\odot}$ model of WW95.
Indices denote the time sequence as explained in the text.}
\label{fig:density}
\end{figure}

\begin{figure}
\begin{center}
\epsfig{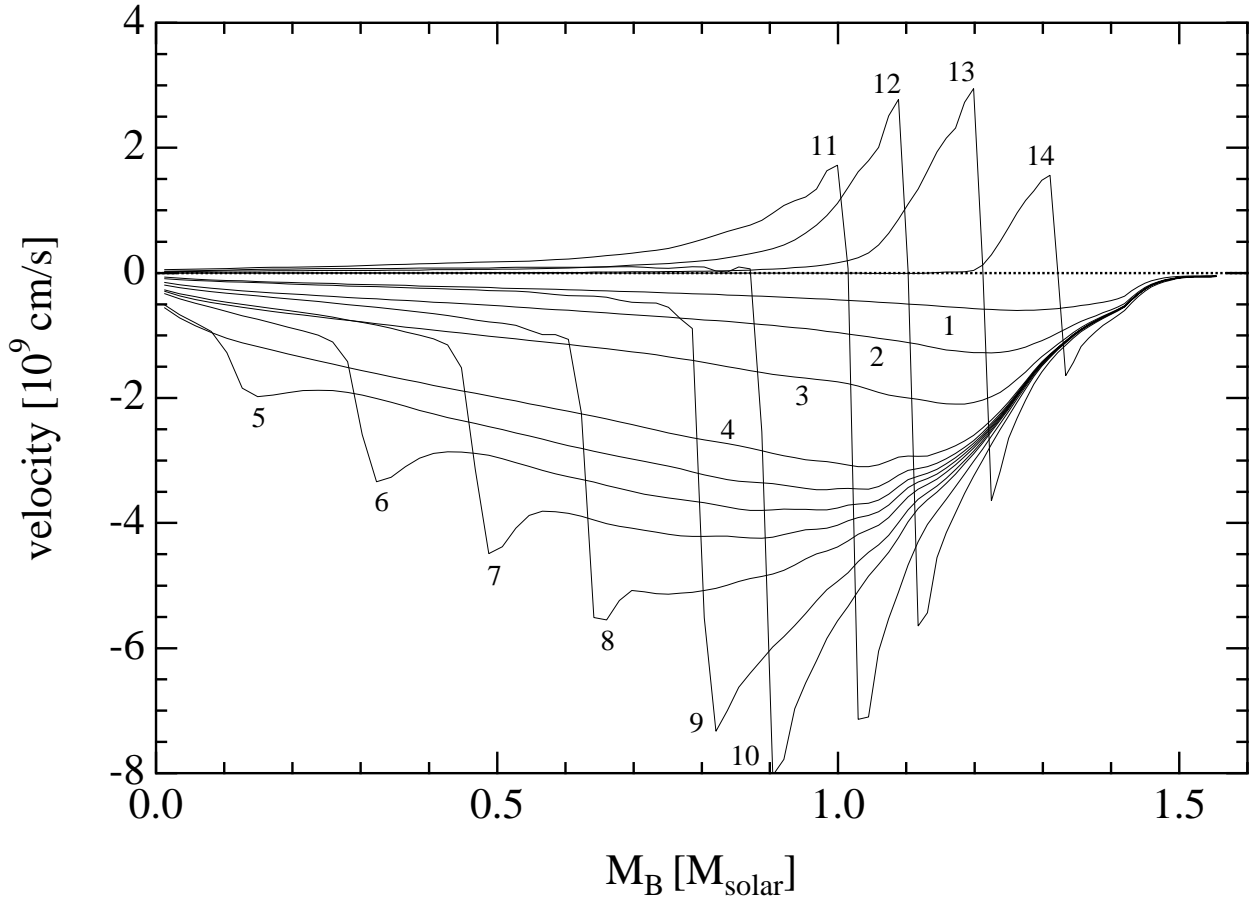}
\end{center}
\caption{The velocity profiles at selected times are shown as a function 
of baryon mass coordinate.}
\label{fig:velocity}
\end{figure}

\begin{figure}
\begin{center}
\epsfig{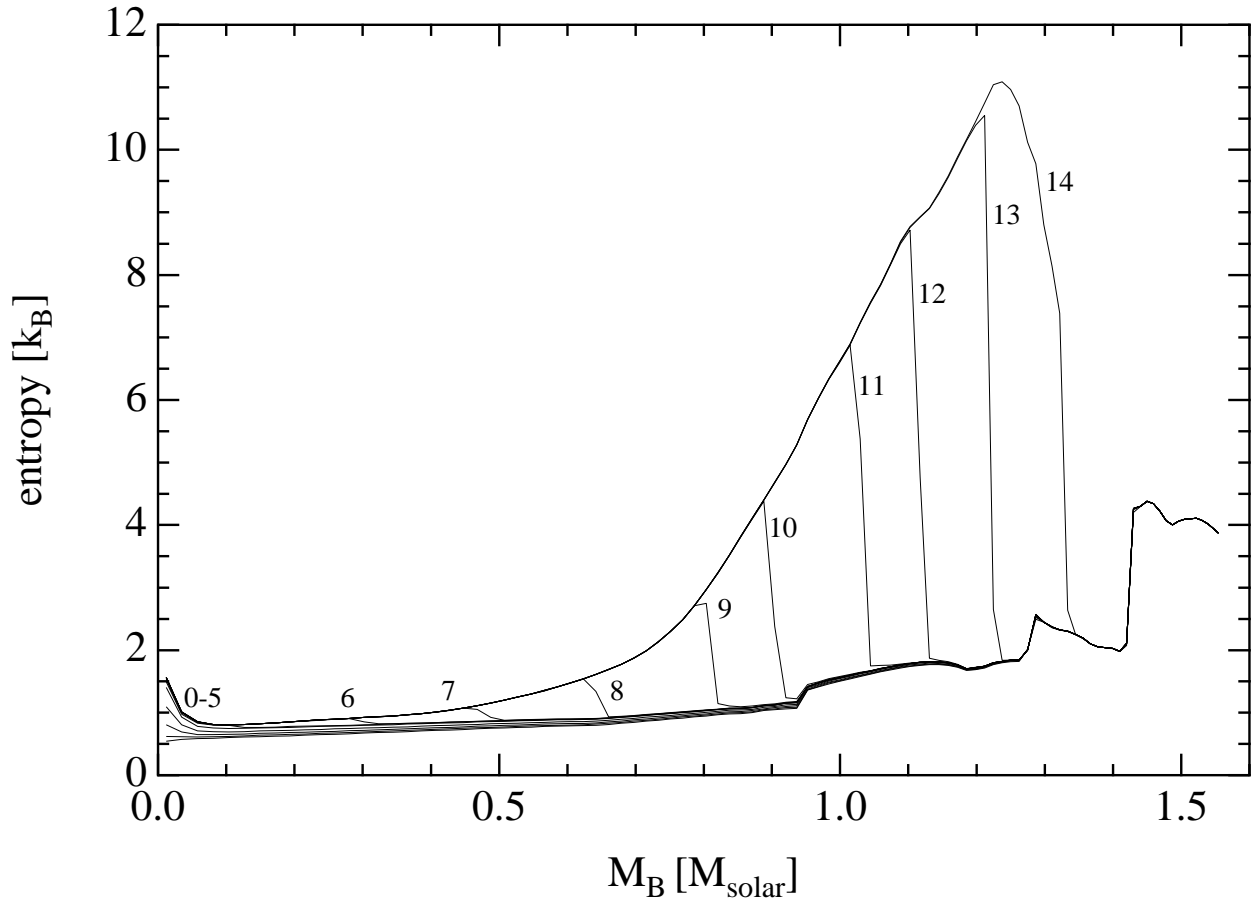}
\end{center}
\caption{The entropy per baryon profiles at selected times are shown as a function 
of baryon mass coordinate.}
\label{fig:entropy}
\end{figure}

It is instructive to display profiles of quantities 
during the core collapse and bounce.  
We show, in Figs. \ref{fig:density}--\ref{fig:entropy}, 
the density, velocity and entropy 
as a function of baryon mass coordinate for various times.  
We choose the case of 15 $M_{\odot}$ model of WW95 
as a representative model and concentrate on the profiles 
of quantities calculated with this model for further 
discussions.

In Fig. \ref{fig:density}, we display the density profiles 
when the central densities reach 
$10^{11}$, $10^{12}$, $10^{13}$, $10^{14}$ g/cm$^{3}$ during 
the collapse (1--4) as well as the density profiles around and 
after the bounce (5--14).  
The initial density profile is also shown (0).
The central density becomes high 
($2.1 \times 10^{14}$ g/cm$^{3}$) at the stage (5) 
and a density jump propagates outward.  
The peak central density is $4.0 \times 10^{14}$ g/cm$^{3}$ 
at the stage (10), which corresponds to the bounce 
at the time 322 msec after the onset of the collapse. 
The central density decreases slightly to 
$3.3 \times 10^{14}$ g/cm$^{3}$ at the stage (14) 
at the time 12 msec after the bounce.  

In the velocity profiles in Fig. \ref{fig:velocity}, 
the development of shock structure is apparent.  
At the stages (1--4), the inner homologous collapse and 
the outer free fall can be seen.
A pressure wave starts at the center (5) and 
develops at the stages (6--9).  
It becomes a shock wave at the border of 
the inner core at the stage (10).  
The positive velocity peak starts growing and 
the outward shock wave proceeds toward the surface of the 
iron core (11--14).  

The profiles of entropy per baryon are shown in Fig. \ref{fig:entropy}.
Because it is an adiabatic collapse, the central entropy 
per baryon stays around $1 k_{B}$ (1--5).  
The entropy is raised as the shock wave propagates and 
reaches above a few $k_{B}$ at the bounce (10).  
The shocked envelope with high entropy ($\sim$ 10 $k_{B}$) 
is formed at (11--14).  
During these sequences, 
the central temperature increases to $\sim$ 5 MeV during 
the collapse (1--5) and becomes as high as 10 MeV 
at the bounce (6--10).  
The temperature peak is formed at the outer core 
due to the shock passage and the temperature exceeds 
20 MeV at the maximum (11--14).  

We describe here shortly a numerical simulation with 
the Lattimer-Swesty EOS of which results are also 
shown for comparison.  
We run a hydrodynamical calculation 
starting from the same initial configuration 
of the 15 $M_{\odot}$ model of WW95.  
A prompt explosion occurs with the explosion energy 
of 1.9 foe, which is larger than the case of the 
relativistic EOS.  
The peak central density is found to be 
$5.3 \times 10^{14}$ g/cm$^{3}$, which is also higher.  
Using the result of hydrodynamics with the Lattimer-Swesty EOS, 
we compare the compositions at the times 
having the same central densities during the collapse 
and at the times 
having the same position of the shock wave after the bounce.  

\begin{figure}
\begin{center}
\epsfig{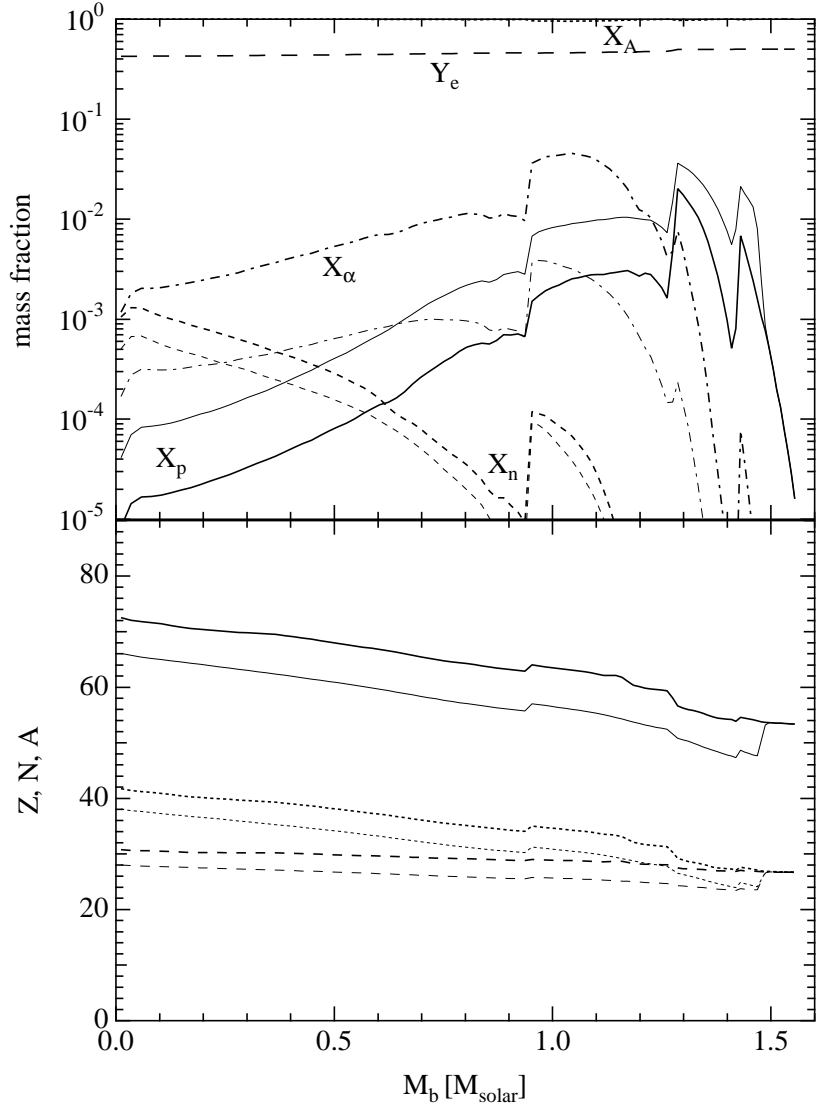}
\end{center}
\caption{The mass fraction of particles (upper panel), 
the mass, neutron and proton numbers of nuclei (lower panel) 
in the core of 15 $M_{\odot}$ model (WW95) are shown 
as a function of baryon mass coordinate.  
The results obtained with the relativistic EOS
are shown by thick lines and the Lattimer-Swesty EOS 
by thin lines.  
Solid, dashed, dotted and dot-dashed lines in the upper panel 
show mass fractions of protons, neutrons, nuclei and alpha 
particles, respectively.
Long dashed line in the upper panel 
displays the initial electron fraction.  
Solid, dotted and dashed lines in the lower panel show 
the mass, neutron and proton numbers of nuclei, respectively.}
\label{fig:initial}
\end{figure}

\begin{figure}
\begin{center}
\epsfig{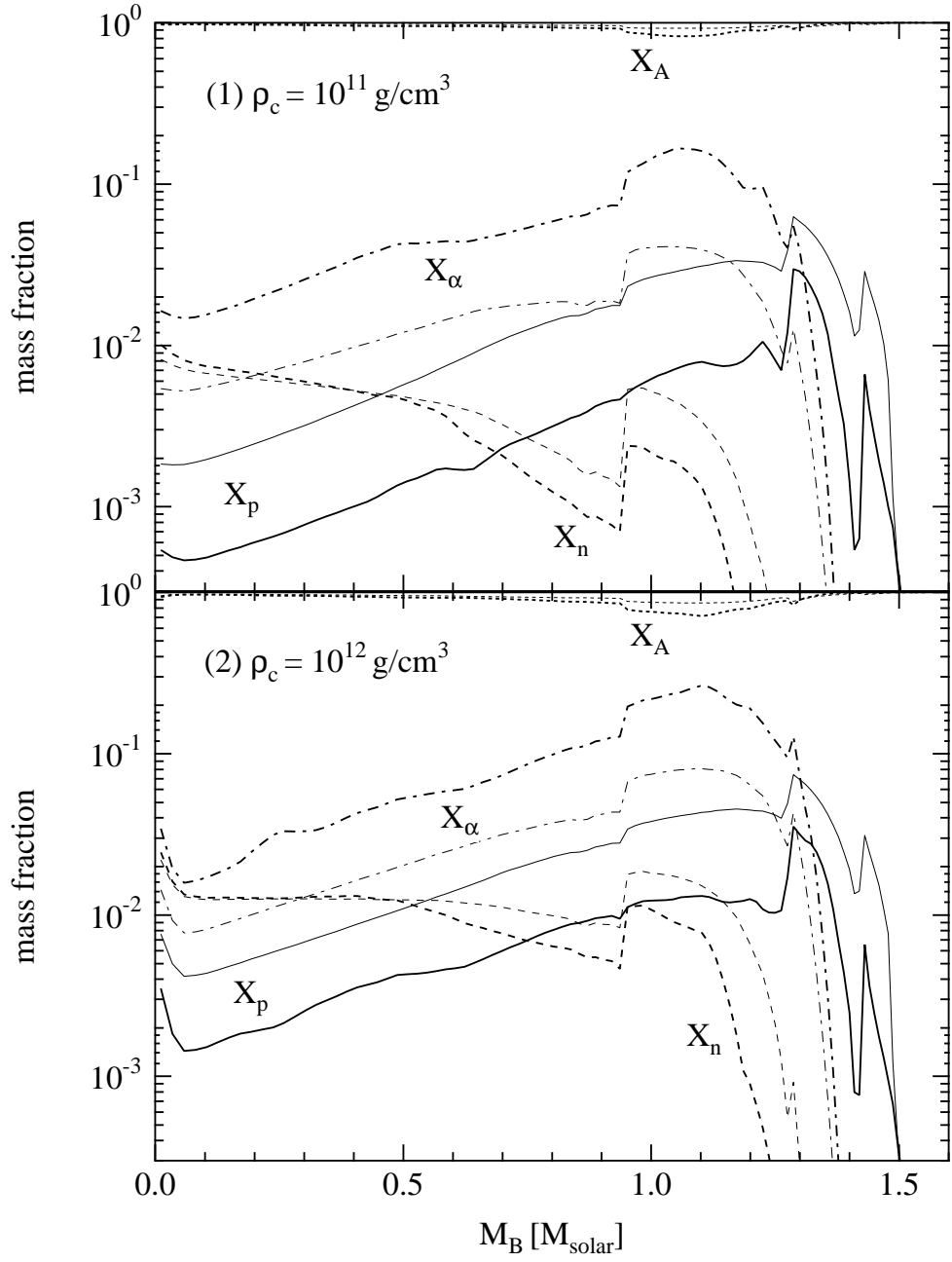}
\end{center}
\caption{The mass fractions of protons, neutrons, nuclei and alpha 
particles are shown as a function of baryon mass coordinate 
at the collapse stages (1, 2) for the case of 15 $M_{\odot}$ model of WW95.  
The notations are the same as in Fig. \ref{fig:initial}.}
\label{fig:composition1}
\end{figure}

\begin{figure}
\begin{center}
\epsfig{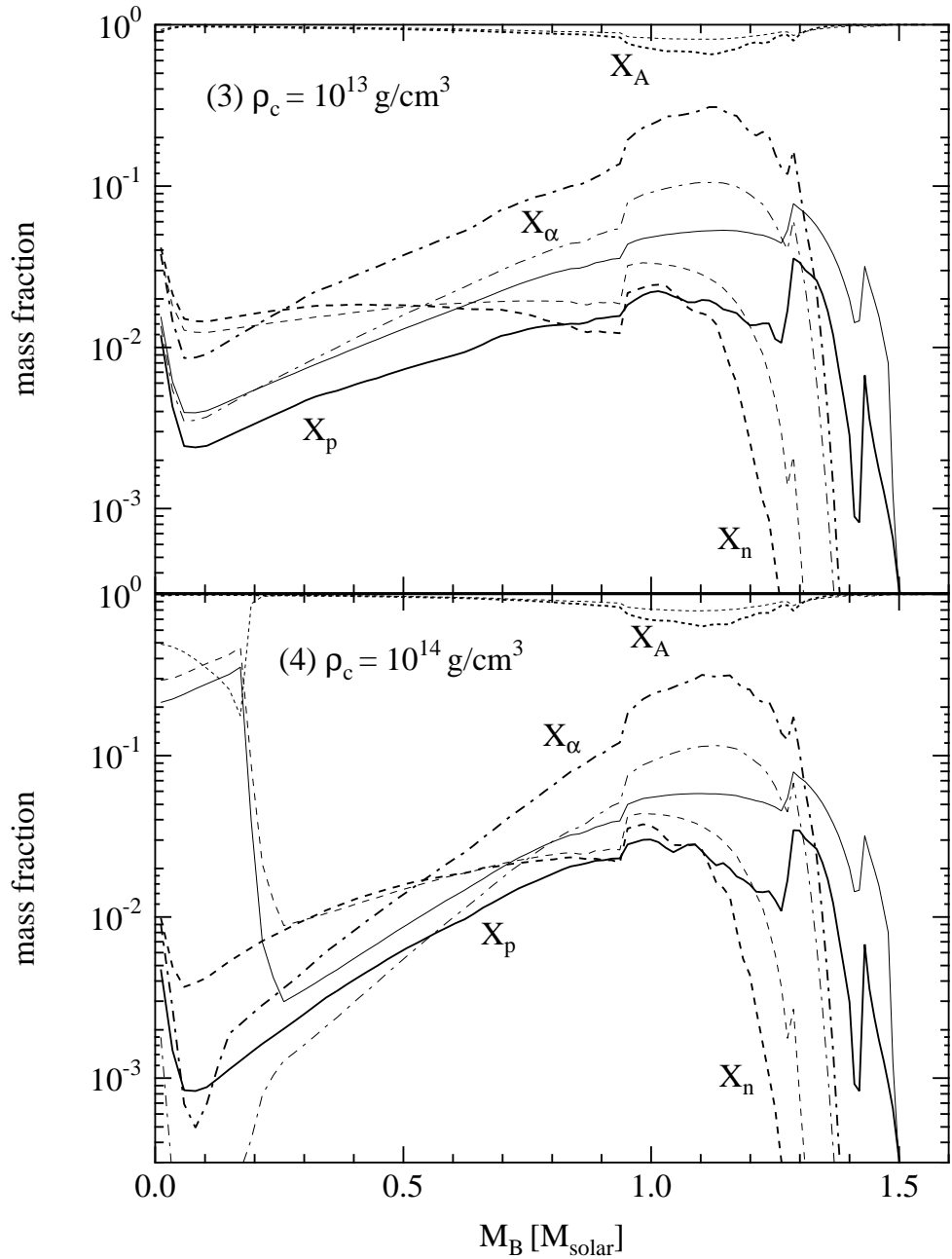}
\end{center}
\caption{The mass fractions of protons, neutrons, nuclei and alpha 
particles are shown as a function of baryon mass coordinate 
at the collapse stages (3, 4).
The notations are the same as in Fig. \ref{fig:initial}.}
\label{fig:composition2}
\end{figure}

\begin{figure}
\begin{center}
\epsfig{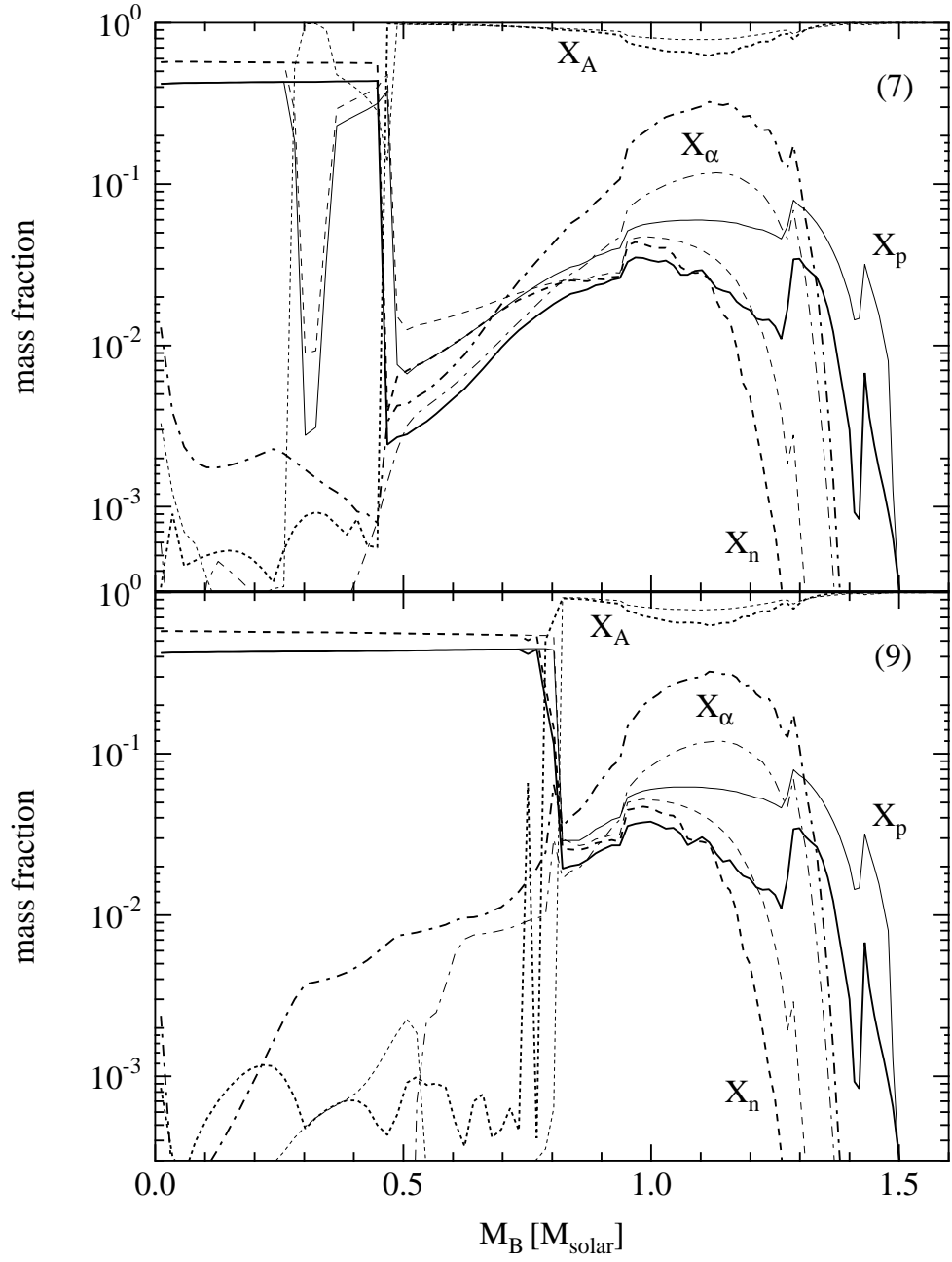}
\end{center}
\caption{The mass fractions of protons, neutrons, nuclei and alpha 
particles are shown as a function of baryon mass coordinate 
at around the bounce stages (7, 9).
The notations are the same as in Fig. \ref{fig:initial}.}
\label{fig:composition3}
\end{figure}

\begin{figure}
\begin{center}
\epsfig{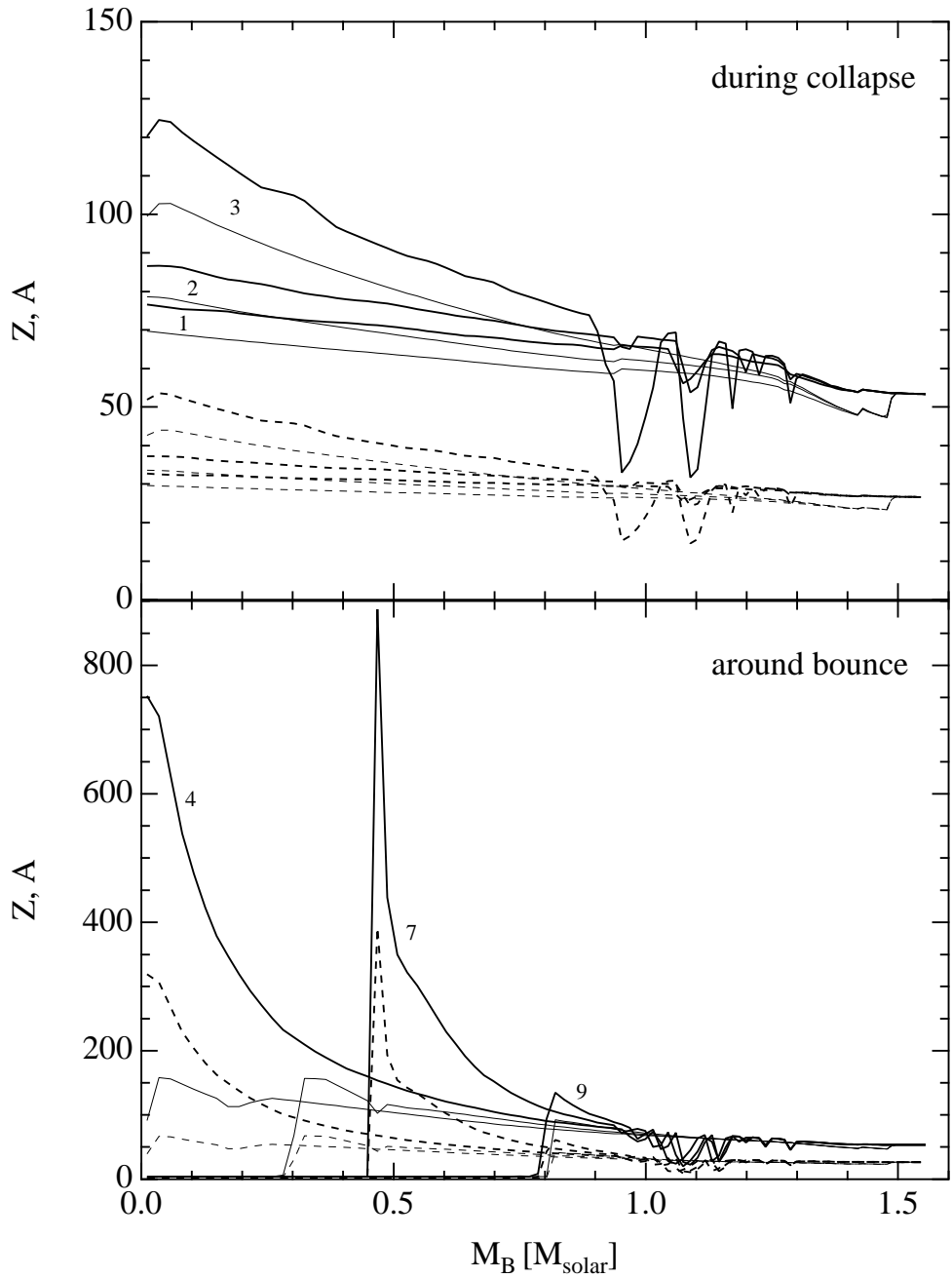}
\end{center}
\caption{The mass and proton number of nuclei 
are shown as a function of baryon mass coordinate 
at the collapse stages (1,2,3) and around the bounce stages (4,7,9).
The notations are the same as in Fig. \ref{fig:initial}.}
\label{fig:ZA}
\end{figure}

The compositions during collapse and bounce are shown 
in Figs. \ref{fig:composition1}--\ref{fig:ZA}.  
The notations are the same as in Fig. \ref{fig:initial}.
The mass fractions during collapse 
are shown as a function of baryon mass coordinate 
at the times (1) and (2) 
(central densities: $10^{11}$, $10^{12}$ g/cm$^{3}$) 
in Fig. \ref{fig:composition1}, 
at the times (3) and (4) ($10^{13}$, $10^{14}$ g/cm$^{3}$) 
in Fig. \ref{fig:composition2}.  
During collapse (1--2) in Fig. \ref{fig:composition1} 
the mass fraction of nuclei decreases slightly 
and alpha particles, protons and neutrons become 
more abundant.  
The difference between the Lattimer-Swesty EOS and 
the relativistic EOS, which is seen in Fig. \ref{fig:initial}, 
is present in early stages of collapse.  
The difference becomes less drastic 
and the compositions become similar in later stages 
(3--4) in Fig. \ref{fig:composition2}.
The mass fraction of alpha particle remains larger 
than that in the Lattimer-Swesty EOS.
Alpha particles are abundant in the outer part 
of the core ($M_{B} > 1.0M_{\odot}$) during core 
collapse and bounce.  
The mass fractions around bounce (7) and (9) are shown 
in Fig. \ref{fig:composition3}.
Neutrons and protons are dominant at the central region 
after bounce because of high density and temperature.  
This region grows due to dissociation of nuclei 
as the shock wave proceeds outward.  
The difference between the relativistic EOS and 
the Lattimer-Swesty EOS in the central region 
is small.  
The appearance of nuclei at $M_{B} \sim 0.3M_{\odot}$ 
for the Lattimer-Swesty EOS at the stage (7) may be 
attributed to their treatment of the phase boundary.

The mass and proton numbers during collapse (1--3) 
and around bounce (4--9) are shown as a function of 
baryon mass coordinate in Fig. \ref{fig:ZA}.
The nuclei become heavier and more neutron-rich as 
the central density increases.  
This tendency is stronger in the relativistic EOS 
than in the Lattimer-Swesty EOS.  
We note that the neutron skin of nuclei is taken into account 
in the relativistic EOS but not in the Lattimer-Swesty EOS 
which may cause different mass number and neutron-richness.  
After bounce, huge nuclei appear in the relativistic EOS 
at the density around $10^{14}$ g/cm$^{3}$.  
This is in contrast to the case of the Lattimer-Swesty EOS, 
which gives nuclei up to the mass number of 200.  
This is because of the different treatment of nuclear shape 
change and bubble phase just below the nuclear matter density.  
In the relativistic EOS table, those effects are not included 
like Lattimer et al. \cite{Lat85} (LLPR) whereas 
the Lattimer-Swesty EOS takes them into account and finds 
smaller nuclear masses than LLPR.  
The zig-zag behavior of mass number and proton number 
around $M_{B} \sim 1.0M_{\odot}$ 
is due to the fact that the density-temperature trajectories 
go around the phase boundary between nuclei and gas, 
where the mass fractions change rapidly.  
%difference, surface energy

\begin{figure}
\begin{center}
\epsfig{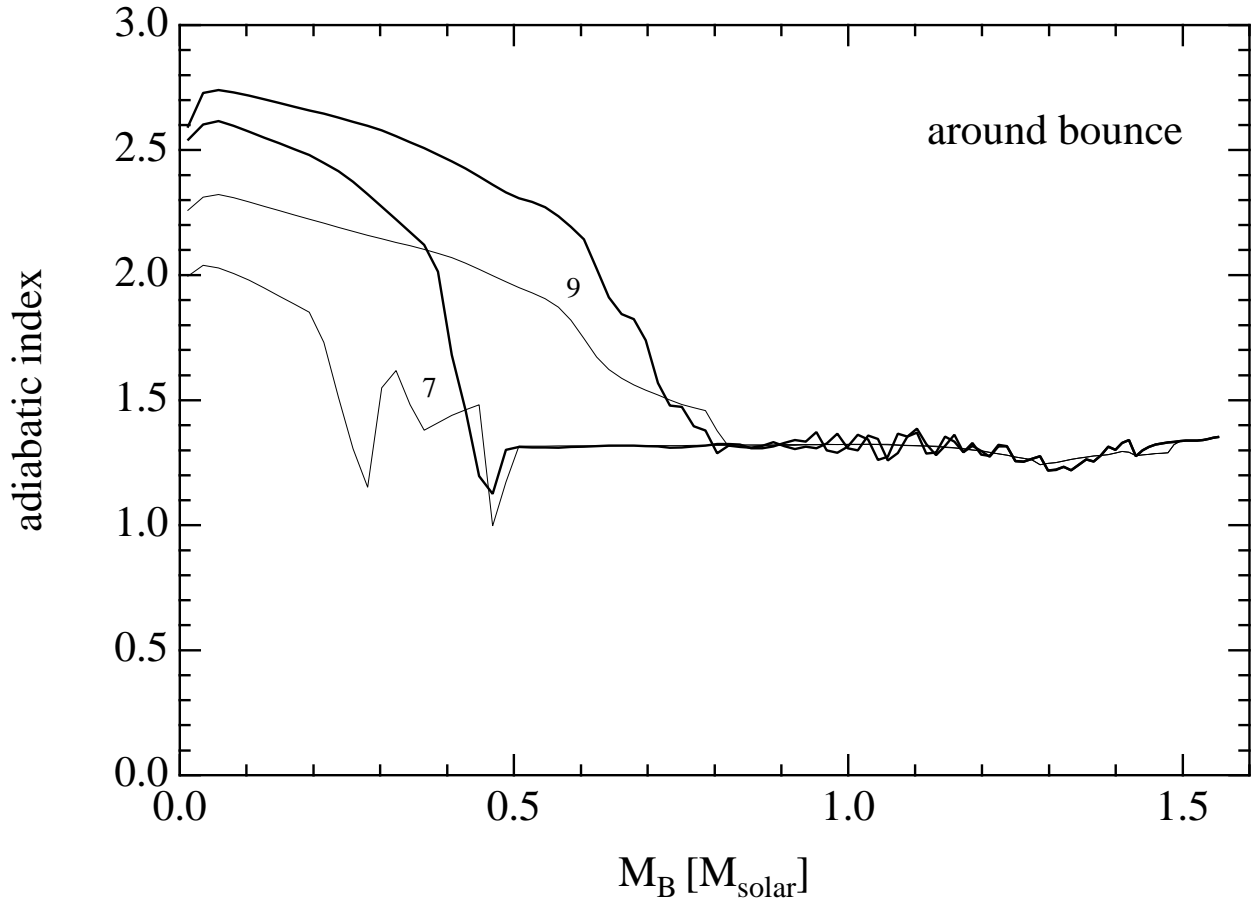}
\end{center}
\caption{The adiabatic indices at around the bounce stages (7,9)
are shown as a function of baryon mass coordinate 
for the relativistic EOS (thick) and the Lattimer-Swesty EOS (thin).}
\label{fig:gamma}
\end{figure}

The adiabatic indices at around the bounce stages (7) and (9) are shown 
as a function of baryon mass coordinate in Fig. \ref{fig:gamma}.
The case of the Lattimer-Swesty EOS is shown by thin lines 
for comparison.  
The adiabatic index in the relativistic EOS 
is larger than that of the Lattimer-Swesty EOS in the central 
region where the density exceeds $10^{14}$ g/cm$^{3}$.  
In the outer region, they are similar to each other except for 
some fluctuations due to the phase transition we mentioned above.  
This difference in the central region indicates that 
the relativistic EOS is stiffer at these densities 
than the Lattimer-Swesty EOS in terms of the adiabatic index.
Since the central density is not far beyond the nuclear matter 
density, the behavior is mostly determined by the incompressibility.  
The value of the incompressibility of the relativistic EOS (K=281 MeV) is 
larger than the value (K=180 MeV) we adopted for the Lattimer-Swesty EOS.

The stiffness can be discussed in terms of the shock energy 
described above and the size of the inner core.  
We have examined whether the explosion energy can be 
explained in terms of the initial shock energy due to the bounce 
of the inner core and the dissociation of iron nuclei 
due to the shock passage in the outer core \cite{Suz94}.  
As a reference, we list the size of inner core in Table 1.  
In Fig. \ref{fig:velocity}, one can clearly see the structure with 
an unshocked inner core and an infalling outer core.  
Here, we define the inner core as the region inside 
the edge of positive velocity 
in the profile when the velocity turns from negative 
to positive 
in the central region (see the bounce stage 10).  
For both WW95 and N97 models, the inner core mass is 
about 0.9 $M_{\odot}$ independent of the progenitor mass.  
The inner cores and remnant masses in the case of 
the Lattimer-Swesty EOS are also similar.  

We plot in Fig. \ref{fig:energy} the profiles of the enclosed energy 
(i.e. $(m_{g}-m_{b})$ in Appendix) 
in the case of 15 $M_{\odot}$ model of WW95.  
The upper and lower panels correspond to 
the stages (10) and (14), respectively.  
The upper panel (a) corresponds to the bounce 
when the shock wave with positive velocity just starts 
to show up and 
the lower panel (b) shows the situation after the 
shock wave have passed through the iron core.  

While the enclosed energy of the inner core 
becomes large at first due to the 
compression and amounts to $1.1 \times 10^{52}$ erg 
at $M_{B}=0.87M_{\odot}$ in Fig. \ref{fig:energy}a, 
it decreases, in Fig. \ref{fig:energy}b, to $5.7 \times 10^{51}$ erg 
as it approaches the final hydrostatic configuration 
due to decompression (see also Fig. \ref{fig:density}).  
The difference of these energies is the initial shock 
energy of $5.6 \times 10^{51}$ erg.  
The shock loses its energy due to the dissociation 
of irons when it passes through the outer iron core.  
The dissociation of irons between $M_{B}=0.87M_{\odot}$ 
(the inner core mass, $M_{inner}$) and $M_{B}=1.14M_{\odot}$ 
(the remnant mass, $M_{remn}$)
requires the energy of $4.3 \times 10^{51}$ erg.  
Due to this energy loss, 
the shock energy is reduced to $1.3 \times 10^{51}$ erg, 
which corresponds to the final explosion energy 
transferred to the ejected material.
In realistic simulations with electron captures, 
the inner core mass will be smaller than the present case 
in which the electron capture and the neutrino interaction 
are artificially suppressed.  
If the inner core mass were smaller by $\sim 0.1M_{\odot}$ 
than the present case, the shock wave would lose 
more energy ($\sim 1.6 \times 10^{51}$ erg, corresponding 
to $\sim 0.1M_{\odot}$) during the propagation 
and, therefore, the prompt explosion should become more difficult.  

We remark that the numerical data table of the relativistic 
EOS works properly in the hydrodynamical calculations 
without troubles 
such as overflows of range due to the lack 
of data or the inconsistency of thermodynamical properties.  
We have monitored the total energy during the numerical 
simulation and have checked that the total energy 
after the hydrodynamical calculation agrees with 
the initial total energy with enough numerical accuracy
of $\sim 10^{-6}$.  

%%%%%%%%%%%%%%%%%%%%%%%%%%%%%%%%%%%%%%%%%%%%%%%%%%%%%%%%%%%%
\section{Outlook on further effects}

It has been discussed that the delayed explosion through 
the revival of shock due to neutrino-heating might be a 
general mechanism of supernova explosion \cite{Jan01}.  
The detailed processes in micro- and macro-physics 
can play significant roles to realize the explosion 
of this type.  
One has to treat carefully all interactions, which change lepton 
fractions in dense matter and lead to heating and cooling.  
We stress that the EOS of dense matter provides the 
information such as compositions and chemical potentials 
and is essential to determine those reaction rates.  
For example, the effective mass of nucleon is tabulated 
in the table.  
It might affect the neutrino opacity substantially \cite{Yam00}.  
Detailed studies by the numerical simulations with 
neutrino transfer using the relativistic EOS table are 
under way \cite{Sum02}.

We have to remark also that the current study is done 
under the spherical symmetry.  
Deviations from the spherical symmetry such as matter mixing 
have been suggested by the observations of light 
curve and line profiles of SN1987A.  
Multi-dimensional effects such as convection 
can play an important role in the explosion mechanism 
\cite{Bet90, Jan01, Eps79}.
However, where, when and how long hydrodynamical instabilities occur 
in the central region of a supernova and how it contributes 
to the explosion mechanism are a matter of long debates.  
On one hand, the convection has been shown to be 
essential for the successful delayed explosion through 
neutron fingers \cite{Wil93}.
On the other hand, it has been claimed that the neutron 
finger convection is unlikely to occur 
\cite{Bru95}.
Other recent studies have demonstrated that the whole 
proto-neutron star just born 
may be convective, which will be favorable for the explosion 
\cite{Kei96} (see, however, \cite{Bur03}).
The diffusive processes in the convection 
have been also discussed recently in detail 
\cite{Mir00}.

In the current study, we have aimed to see the 
properties of the 
relativistic EOS as a basis for further studies.  
Multi-dimensional simulations of supernova explosion 
with the relativistic EOS table is of great interest 
given the present results.  
In order to assess the influence of the relativistic EOS on 
multi-dimensional hydrodynamics, 
we focus on the convective instability 
inside the neutrinosphere and investigate 
the condition derived by the relativistic EOS.

The negative gradients of the entropy and/or the lepton 
fraction are often seen in a number of numerical simulations 
(see \cite{Sum95c} for example) 
and can lead to the convective overturn.  
The convection can raise the hot and lepton-rich matter 
toward the neutrinosphere 
and enhance the neutrino luminosities to heat the region 
just behind the stalled shock.  
This extra heating 
can be crucial to revive the stalled shock 
leading to a successful delayed explosion.  

Here, we discuss a role of EOS in the Ledoux criterion, 
the condition that the convection occurs without diffusive 
processes by the gradient of entropy and lepton fraction.  
This means that the EOS is also essential 
to judge whether convection occurs or not through the 
distributions of these quantities.  
The Ledoux criterion is expressed by,
\begin{equation}
\left(\frac{\partial \rho}{\partial Y_{l}} \right)_{{\rm P},\, S}
\cdot 
\left(\frac{d Y_{l}}{d r} \right) +
\left(\frac{\partial \rho}{\partial S}     \right)_{{\rm P},\, Y_{l}}
\cdot 
\left(\frac{d S    }{d r} \right) \geq 0,
\end{equation}
where $\rho$, $Y_{l}$, $S$ are density, lepton fraction 
and entropy per baryon, respectively.  
The derivatives of density with respect to lepton 
fraction,
\begin{equation}
\left(\frac{\partial \rho}{\partial Y_{l}} \right)_{{\rm P},\, S},
\label{eqn:deriv}
\end{equation}
and with respect to entropy,
\begin{equation}
\left(\frac{\partial \rho}{\partial S}     \right)_{{\rm P},\, Y_{l}},
\end{equation}
are usually assumed to be negative.  
Hence the negative gradient of entropy 
and/or lepton fraction inside stars leads to the convection.  
However, the derivative of density with respect to lepton 
fraction can change its sign in a certain environment \cite{Lat81}.  
Note that the derivative of density with respect to entropy 
per baryon is always negative.  

\begin{figure}
\begin{center}
\epsfig{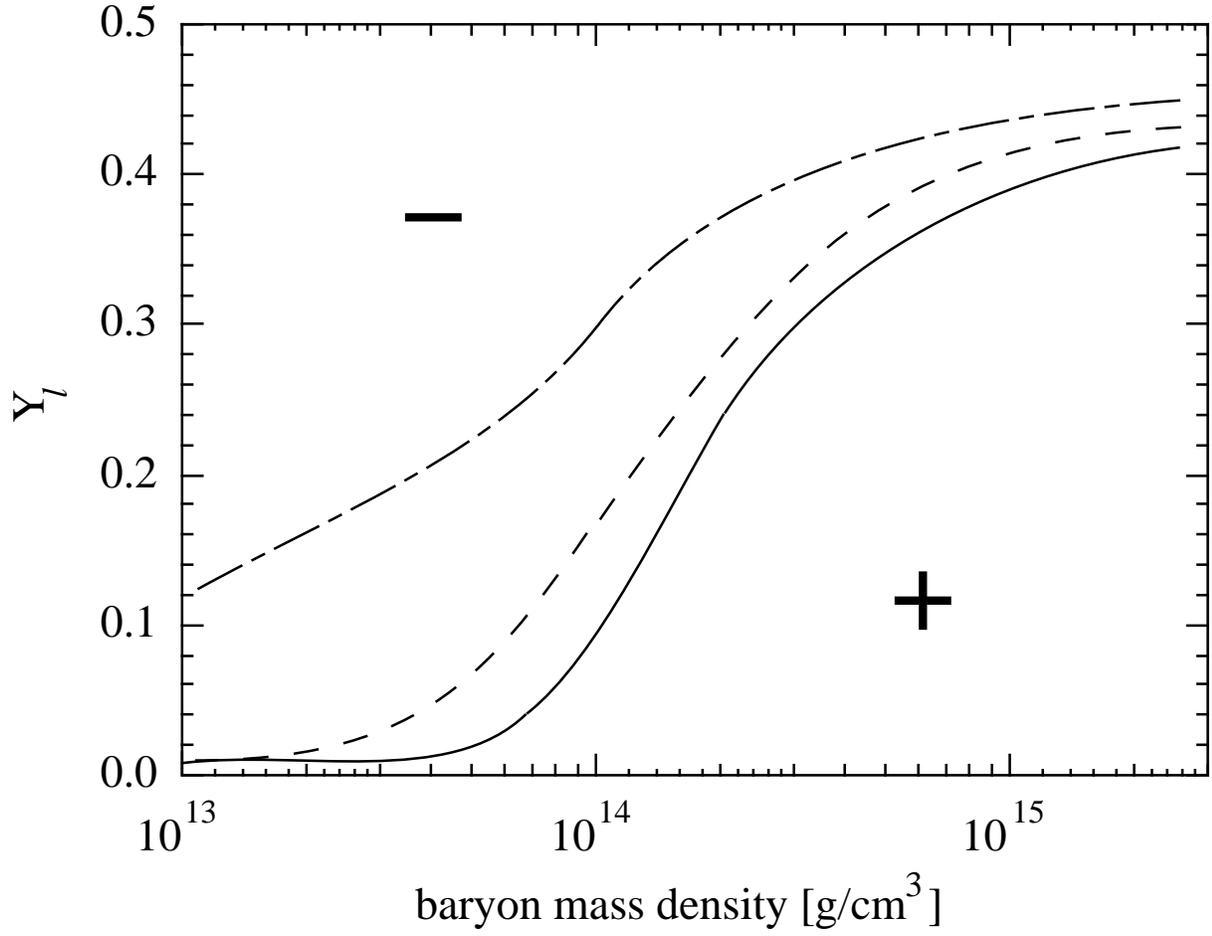}
\end{center}
\caption{The sign of derivative of thermodynamical quantities, which 
determine convective condition, is displayed in density-lepton 
fraction plane.  See the definition in text.}
\label{fig:convection}
\end{figure}

We investigate this point with the relativistic EOS.  
For this purpose, we consider the dense matter 
under thermal and beta equilibrium among neutrons, protons, 
electrons, positrons, three flavor (anti-)neutrinos and photons.
Basic properties of this supernova matter have been reported 
in \cite{Sum95b}.  
Figure \ref{fig:convection} displays 
the sign of derivative and the borders of the 
sign change for the cases of the entropy per baryon $S=1$, 2 and 4 
by solid, dashed and dash-dotted lines, respectively.  
It is remarkable that the derivative is positive in the 
region with high density and low lepton fraction.  
This is in accord with the previous results by Lattimer 
and Mazurek \cite{Lat81} and other recent works 
\cite{Bru96, Mez98}.  
We note, however, that the positive sign region tends 
to be much wider than the previous results at high densities 
due to the effect of the larger symmetry energy.  
We stress here an important role of the nuclear symmetry energy, 
which is well constrained by the experiments of unstable nuclei, 
and is properly taken into account in the relativistic EOS.  

The sign change happens 
because the nuclear interaction contributes 
to the pressure on top of the lepton contribution.  
The contribution of symmetry energy dominates in the neutron-rich 
(low lepton fraction) environment.  
An increase in lepton fraction not only leads to 
the increase in lepton pressure, but also leads to 
the decrease of baryonic pressure due to nuclear interactions.  
Therefore, the 
density must increase to keep the pressure constant in this positive 
sign regime.  
The convection is then stabilized 
by the negative gradient of lepton fraction.  
On the contrary, the positive gradient of lepton fraction 
can lead to convection if the entropy per baryon is constant.  
Therefore, an extra attention should be paid to the convective region 
itself in multi-dimensional simulations 
with the relativistic EOS.  

%%%%%%%%%%%%%%%%%%%%%%%%%%%%%%%%%%%%%%%%%%%%%%%%%%%%%%%%%%%%
\section{Summary}

We have studied the properties of the relativistic EOS 
during the core collapse and bounce in supernovae.  
The relativistic EOS table is newly derived based on the relativistic 
many body theory checked by the experimental data of unstable 
nuclei and is available for astrophysical simulations.  
To examine the properties of dense matter given by the 
relativistic EOS in dynamical 
situations of supernova explosion, we have utilized the 
general relativistic hydrodynamical calculations of 
gravitational collapse of iron cores.  

In order to provide the basic information of the relativistic 
EOS before we proceed to detailed numerical simulations,
we have followed the adiabatic collapse with the fixed 
electron fraction without neutrino transfer and have obtained 
model explosions.  
Because of the high electron fraction, which is fixed at 
the initial value, the inner core mass 
is large and the hydrodynamical explosion is obtained.
We have examined the thermodynamical quantities and 
compositions of dense matter during collapse and bounce 
in these model explosions.  
We have compared the compositions with those 
in the Lattimer-Swesty EOS by performing the corresponding 
adiabatic collapse calculation.  
We have evaluated the explosion energy 
for a series of progenitor models as one of measures of 
the stiffness of EOS.  
We have also discussed that the relativistic EOS 
might have an influence on the condition of convection.  
We have seen that 
the numerical data table of the relativistic EOS 
works quite successfully in the numerical simulations.  

We have found that the initial composition in presupernova 
cores is fairly different from that in the Lattimer-Swesty EOS.  
The difference of composition persists during early stages of collapse 
and becomes less significant at and after bounce.  
The mass fraction of protons at the early stages is much smaller 
than that in the Lattimer-Swesty EOS 
because of the large symmetry energy in the relativistic nuclear
many body theory.  
The mass fraction of alpha particles is 
larger during the whole stage.  
The species of nuclei is found to be heavier and more neutron-rich 
during collapse and the mass number turns out to be quite large 
after bounce.  
Compositional differences may lead to changes in electron capture 
rates and neutrino interaction rates and might influence 
the dynamics of supernova explosion.  

These differences of compositions should be examined further 
in numerical simulations of hydrodynamics with neutrino transfer 
by switching the two sets of EOS.  
Detailed simulations including the neutrino transfer 
with this new relativistic EOS table are called for 
to examine the role of the EOS in the delicate mechanism 
of supernova explosion.  
Such efforts of numerical simulations are currently 
being made \cite{Sum02}.

%%%%%%%%%%%%%%%%%%%%%%%%%%%%%%%%%%%%%%%%%%%%%
\section*{Acknowledgment}

We would like to express special thanks to Hong Shen 
for her devoted achievement of the relativistic EOS table 
and to H. Ono for his great efforts on the tabulation of 
the Lattimer-Swesty EOS.
We would like to thank Stan Woosley and Ken'ichi Nomoto 
for providing us with 
their numerical data of presupernova models.  
K. S. is grateful to M. Terasawa, G. Mathews, 
T. Kajino and I. Tanihata for fruitful collaborations 
on r-process nucleosynthesis in supernova explosions.  
K. S. also thanks S. Wanajo, A. Onishi, K. Oyamatsu 
and Thomas Janka for stimulating discussions.  
The numerical simulations have been performed on the 
supercomputers at RIKEN and 
KEK (KEK Supercomputer Projects No.01-75 and No.02-87).  
This work is partially supported by the Grants-in-Aid for the
Center-of-Excellence (COE) Research of the ministry of Education,
Science, Sports and Culture of Japan to RESCEU (No.07CE2002).
This work is also supported in part 
by Japan Society for Promotion of Science, and by the Grant-in Aid for
Scientific Research (12047230, 12740138, 13740165, 14039210, 14740166, 
15740160)
of the Ministry of Education, Science, Sports and Culture of Japan.

%%%%%%%%%%%%%%%%%%%%%%%%%%%%%%%%%%%%%%%%%%%%%
\section*{Appendix}
\setcounter{equation}{0}
\renewcommand{\theequation}{A-\arabic{equation}}
We summarize here the definitions of gravitational mass, 
baryon mass and explosion energy in general relativity.  
The notation mostly follows those in the reference \cite{Yam97}.
We take $c = G = 1$ in the following equations.  

The baryon mass elements in spherical symmetry is given by
\begin{equation}
dm_{b}=             \frac {\rho_{b}}{\Gamma} \, 
4 \pi r^{2} dr,
\end{equation}
and the baryon mass coordinate is defined accordingly as 
\begin{equation}
 m_{b} =  \int_{0}^{r} \frac {\rho_{b}}{\Gamma} 
             4 \pi r^{\prime \, 2} dr^{\prime}.
\end{equation}
Here, $\rho_{b}$ is the baryon mass density and 
$\Gamma$ is the relativistic gamma factor defined by, 
\begin{equation}
\Gamma^{2} = 1 + U^{2} - \frac{2 m_{g}}{r}.
\end{equation}
$U$ is the radial fluid velocity and $m_{g}$ is 
the gravitational mass defined below.  
The gravitational mass, $dm_{g}$, of the baryon mass elements, 
$dm_{b}$, is given by 
\begin{eqnarray}
dm_{g} & = & \rho_{b} \, (1 + \varepsilon ) \, 4 \pi r^{2} dr, \\
       & = & \Gamma (1 + \varepsilon ) dm_{b},
\end{eqnarray}
where $\varepsilon$ is the specific internal energy density.  
The gravitational mass is then given by 
\begin{eqnarray}
m_{g} & = & \int_{0}^{m_{b}} \frac{dm_{g}}{dm_{b}} \, dm_{b}, \\
%      & = & \int_{0}^{m_{b}} \Gamma (1 + \varepsilon ) \, dm_{b}, \\
      & = & \int_{0}^{r} \rho_{b} \, (1 + \varepsilon ) \, 
            4 \pi r^{\prime \, 2} dr^{\prime}. 
\end{eqnarray}
The gravitational mass of ejecta is evaluated by
\begin{equation}
M_{g}^{ejecta}=\int_{M_{remn}}^{M_{calc}}\frac{dm_{g}}{dm_{b}} \, dm_{b},
\end{equation}
where the range of integral in baryon mass coordinate 
is from the remnant mass, $M_{remn}$, to the total baryon 
mass covered in the calculation, $M_{calc}$.  
The remnant mass is defined at the position in the baryon mass 
coordinate where the minimum of the enclosed energy ($m_g - m_b$)
occurs (see Fig. \ref{fig:energy}, for example).
The explosion energy is calculated as 
\begin{equation}
E_{exp}=M_{g}^{ejecta} - M_{b}^{ejecta},
\end{equation}
where the baryon mass of ejecta is 
\begin{equation}
M_{b}^{ejecta}=M_{calc} - M_{remn}.  
\end{equation}
\newpage

\section*{Table caption}

\begin{description}

\item[Table 1]
The list of models with different progenitor masses 
($M_{proj}$) of WW95 (a) and N97 (b).  
The sizes (in baryon mass) of iron core ($M_{Fe}$), core covered in 
calculation ($M_{calc}$) and inner core ($M_{inner}$) 
obtained in simulations are listed.  
The remnant mass ($M_{remn}$) and explosion energy ($E_{exp}$) are listed 
for successful cases.  
\end{description}

%%%%%%%%%%%%%%%%%%%%%%%%%%%%%%%%%%%%%%%%%%%%%%%%%%%%%%%%%%%%%%%%%%%%
%\newpage

\begin{center} {\bf Table 1a}
\end{center}
\[\begin{tabular}{ccccccc} \hline
Model & $M_{proj}$ [$M_{\odot}$] & $M_{Fe}$ [$M_{\odot}$] & $M_{calc}$ [$M_{\odot}$] & $M_{inner}$ [$M_{\odot}$] & $M_{remn}$ [$M_{\odot}$] & $E_{exp}$ [$10^{51}$ erg] \\ \hline
S11A & 11 & 1.32 & 1.42 & 0.87 & 1.15 & 1.7 \\
S12A & 12 & 1.32 & 1.36 & 0.86 & 1.08 & 1.7 \\
S13A & 13 & 1.41 & 1.62 & 0.87 & $-$ & $-$ \\
S15A & 15 & 1.32 & 1.56 & 0.87 & 1.14 & 1.5 \\
S18A & 18 & 1.46 & 1.84 & 0.89 & $-$ & $-$ \\
S20A & 20 & 1.74 & 2.20 & 0.88 & $-$ & $-$ \\ \hline
S15A-LS & 15 & 1.32 & 1.56 & 0.87 & 1.15 & 1.9 \\ \hline
\end{tabular} \]

\begin{center} {\bf Table 1b}
\end{center}
\[\begin{tabular}{ccccccc} \hline
Model & $M_{proj}$ [$M_{\odot}$] & $M_{Fe}$ [$M_{\odot}$] & $M_{calc}$ [$M_{\odot}$] & $M_{inner}$ [$M_{\odot}$] & $M_{remn}$ [$M_{\odot}$] & $E_{exp}$ [$10^{51}$ erg] \\ \hline
3.3 & 13 & 1.18 & 1.38 & 0.83 & 1.06 & 1.8 \\
4   & 15 & 1.28 & 1.41 & 0.83 & 1.08 & 1.7 \\
5   & 18 & 1.36 & 1.62 & 0.85 & 1.13 & 1.7 \\
6   & 20 & 1.40 & 1.67 & 0.84 & $-$ & $-$ \\ \hline
\end{tabular} \]

%%%%%%%%%%%%%%%%%%%%%%%%%%%%%%%%%%%%%%%%%%%%%%%%%%%%%%%%%%%%%%%%%%%%
\newpage
%%%%%%%%%%%%%%%%%%%%%%%%%%%%%%%%%%%%%%%%%%%%%%%%%%%%%%%%%%%%%%%%%%%%
%\ref{figure:initial}
%Sumi03/Research/Hydro/collapse3/vw15ms04-ml02/logb.1
%
%Sumi03/Research/Hydro/collapse3/vw15ms04.1/radius
%
%vw15ms04.1
% the step numbers corresponding indices in figures
% 0:    1
% 1:  320
% 2:  550
% 3:  780
% 4: 1010
% 5: 1120
% 6: 1330
% 7: 1630
% 8: 2120
% 9: 2960
%10: 3680
%11: 4990
%12: 6090
%13: 7890
%14:11310
%
%Sumi03/Research/Hydro/collapse3/vw20ms01.1/radius
%
%Sumi03/Research/Hydro/collapse3/vw15ms04.1/vw15ms04
%
%Sumi03/Research/Hydro/collapse3/vw15ms04.1/vw15ms04
%
%\begin{figure}
%\begin{center}
%\epsfig{file=temperature,width=18cm}
%\end{center}
%\caption{The temperature profiles at selected times are shown as a function 
%of baryon mass coordinate.}
%\label{fig:temperature}
%\end{figure}
%Sumi03/Research/Hydro/collapse3/vw15ms04.1/vw15ms04
%
%Sumi03/Research/Hydro/collapse3/vw15ms04.1/vw15ms04
%
%Sumi03/Research/Hydro/collapse3/vw15ms04-ml02/vw15ms04-ml02
%
%vw15ml02.1
% the step numbers corresponding indices in figures
% 0:    1
% 1:  330
% 2:  560
% 3:  790
% 4: 1080
% 5: 
% 6: 
% 7: 1510
% 8: 
% 9: 3130
%10: 
%11: 5010 (not used)
%12: 
%13: 
%14: 
%
%Sumi03/Research/Hydro/collapse3/vw15ms04-ml02/vw15ms04-ml02
%
%Sumi03/Research/Hydro/collapse3/vw15ms04-ml02/vw15ms04-ml02
%
%Sumi03/Research/Hydro/collapse3/vw15ms04-ml02/vw15ms04-ml02
%
%Sumi03/Research/Hydro/collapse3/vw15ms04-ml02/vw15ms04-ml02
%
\begin{figure}
\begin{center}
\epsfig{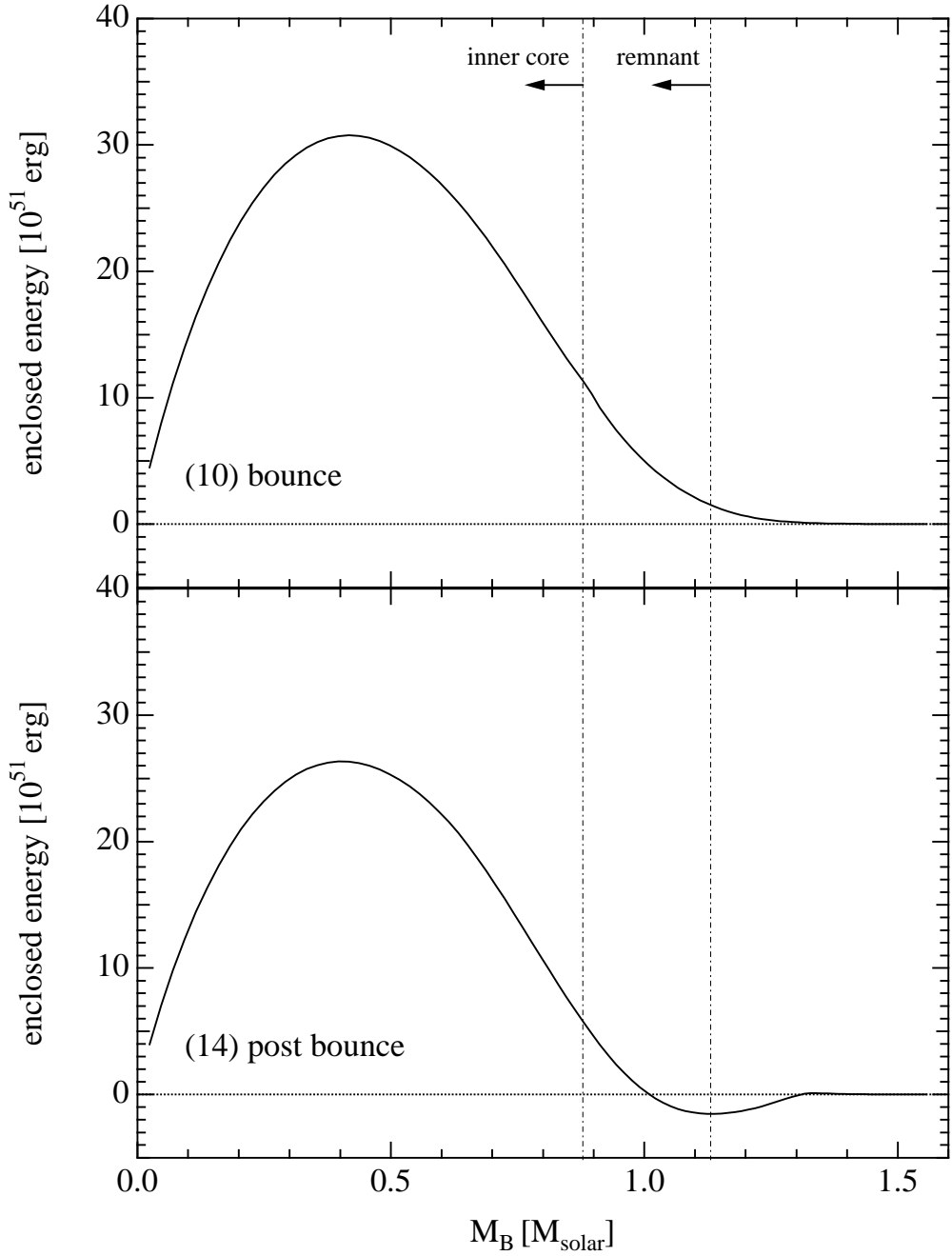}
\end{center}
\caption{The profiles of the enclosed energy are shown 
as a function of baryon mass coordinate 
at the bounce stages (10,14)
%a) at time 367 msec and b) at time 387 msec 
for the case of 15 $M_{\odot}$ model of WW95.}
\label{fig:energy}
\end{figure}
%Sumi03/Research/Hydro/collapse3/vw15ms04.1/eprofile2.03680-11310
%
%\epsfig{file=convection.EPSF}}
%Sumi01/Documents/Papers/Supernova/figures
%
%
% \begin{figure}
% \begin{center}
% \epsfig{file=figure4d.EPSF,width=18cm}
% \end{center}
% \end{figure}
%
%%%%%%%%%%%%%%%%%%%%%%%%%%%%%%%%%%%%%%%%%%%%%%%%%%%%%%%%%%%%%%%%%%%%
\end{document}